\documentclass[a4paper,11pt]{extarticle}  

\usepackage{stylesheet}

\usepackage[left=1in,top=1in,bottom=.65in,right=1in]{geometry}

\usepackage{amssymb}
\usepackage{booktabs}
\usepackage[ruled,vlined, linesnumbered]{algorithm2e}
\usepackage{textcomp}

\title{Bridging the climate to energy data gap: simulated annealing for representative climate year selection}

\author{%
\textbf{%
Bram van Duinen \orcidlink{0000-0003-3683-6304}\textcolor{Accent}{\textsuperscript{1,*}},
Karin van der Wiel \orcidlink{0000-0001-9365-5759}\textcolor{Accent}{\textsuperscript{1,2}},
Jean Thorey{\textsuperscript{3}}, %
Laurens Stoop \orcidlink{0000-0003-2756-5653}\textcolor{Accent}{\textsuperscript{4}} %
}\\[0.5em]
\begin{small}
\textcolor{Accent}{\textsuperscript{1}}Royal Netherlands Meteorological Institute (KNMI), De Bilt, the Netherlands\\[0.5em] 
\textcolor{Accent}{\textsuperscript{2}}Faculty of Geosciences, Utrecht University, Utrecht, the Netherlands\\[0.5em] 
\textcolor{Accent}{\textsuperscript{3}}Réseau de Transport d'Électricité, Paris, France\\[0.5em] 
\textcolor{Accent}{\textsuperscript{4}}TenneT, Arnhem, the Netherlands\\[0.5em] 
\textcolor{Accent}{\textsuperscript{*}}Corresponding Author: \textcolor{Accent}{bram.van.duinen@knmi.nl} \\
\emph{This manuscript has been submitted to Applied Energy for consideration}\end{small}
}

\begin{document}







\pagenumbering{arabic}

\maketitle

\begin{abstract}
Energy system models are increasingly dependent on representative climate input. Yet, a fundamental mismatch persists between the hundreds of simulated years often used in climate science and the handful of years that computationally demanding power system models can process. Current practice, including ENTSO-E's European Resource Adequacy Assessment, relies on climate year selections that have not been validated against explicit representativeness criteria. This risks biased investment decisions and blind spots for plausible weather conditions. This study proposes simulated annealing as an optimisation method for selecting representative subsets of complete climate years from large climate ensembles. Representativeness is quantified using the seasonal sliced Wasserstein distance, a metric from optimal transport theory that captures representativeness on marginal distributions, inter-variable correlations, and seasonal structure simultaneously. We evaluate simulated annealing against the alternative methods random search, filtered random search, and K-Medoids clustering across three test cases spanning the Netherlands and Europe, using 180 climate years from the Pan-European Climate Database as a reference. Simulated annealing consistently produces the most representative subsets and outperforms all compared methods. Simulated annealing achieves an effective sample size four to five times the actual subset size. The resulting subsets are roughly 2.5--3.5 times more representative than current ENTSO-E practice. The method is application-agnostic and its output can serve as a validated climate data input to any subsequent (energy) impact study.
\end{abstract}

\section*{Highlights}
\begin{itemize}
    \item Simulated annealing selects representative climate years for energy system models
    \item Seasonal sliced Wasserstein distance adequately captures representativeness
    \item Simulated annealing outperforms currently used climate year selection methods
    \item Effective sample size reaches four to five times the optimised subset size
    \item Optimised subsets are 2.5-3.5 times more representative than current practice
\end{itemize}

\noindent{\it \color{Highlight} Keywords}: Climate year selection, Simulated annealing, Sliced Wasserstein distance, Energy system modelling, Representative periods, Representative years, Large ensemble modelling

\section{Introduction}
Climate variability is one of the central challenges in energy system planning, but in current practice is often at risk of being averaged away. As electricity systems become increasingly dominated by weather-dependent renewable sources \cite{intergovernmentalpanelonclimatechangeipccEnergySystems2023}, it becomes crucially important to have accurate climate data input \cite{rugglesPlanningReliableWind2024}. For transmission system operators conducting large-scale power system optimisation studies, the choice of climate years used as model input therefore carries substantial weight. Yet this methodological decision has received limited systematic attention. A climate year, or set of climate years, is a representative year/set that shares relevant aspects with the full climate of the study period. This can for example be a single year constructed from long-term observations to reflect average conditions, or a small ensemble of years selected to also capture extremes such as unusually cold winters, low-wind periods, or high solar irradiance. Generally, climate years serve as forcing input into energy system models, and hence their selection determines how well the modelled energy system captures the effects of climate variability. \par

The usage of properly chosen, representative climate years is essential for accurate investment planning and resource adequacy assessments. Unrepresentative and thus biased climate years used as input to energy system studies can easily cost billions of societal funds due to misinformed investments \cite{grochowiczIntersectingNearoptimalSpaces2023, gotskeDesigningSectorcoupledEuropean2024, pecoraQuantifyingImpactsWeather2025}. Moreover, biased climate year selections may systematically exclude weather conditions that climate science deems plausible, leaving the energy system unprepared for events that are not genuine surprises but artefacts of poor methodology \cite{pfenningerDealingMultipleDecades2017, rugglesPlanningReliableWind2024, kelderHowStopBeing2025, pecoraQuantifyingImpactsWeather2025, obergImpactInterannualWeather2025, chatzistylianosAssessingImpactClimate2026}. Yet current practice across energy planning studies, including ENTSO-E's European Resource Adequacy Assessment and Ten-Year Network Development Plan \cite{TYNDP2024Scenarios, EuropeanResourceAdequacy}, relies on selection methods that were adopted without rigorous validation against representativeness criteria. \par

To adequately capture climate variability \cite{suarez-gutierrezInternalVariabilityEuropean2018}, climate scientists have developed large-ensemble modelling approaches. Here, hundreds to thousands of years are simulated to represent the full distribution of possible outcomes in a given climate \cite{vanderwielAddedValueLarge2019,  vanderwielMeteorologicalConditionsLeading2019,bevacquaAdvancingResearchCompound2023, muntjewerfKNMILargeEnsemble2023}. The chaotic nature of the climate system means that subsequent years can turn out completely different, despite having near-identical climatic background state. This makes large sample sizes essential for robust statistical characterisation of both typical conditions and extremes. \par

However, a fundamental mismatch exists between climate science and energy system modelling \cite{bloomfieldQuantifyingIncreasingSensitivity2016, craigOvercomingDisconnectEnergy2022, gotskeDesigningSectorcoupledEuropean2024, rugglesPlanningReliableWind2024, vanduinenMeteorologicalDriversCooccurring2025}. While state-of-the-art climate datasets may span hundreds to thousands of simulated years, detailed industry-standard power system optimisation models are computationally demanding and cannot handle such large input datasets. These models must optimise processes such as unit commitment, network constraints, and market dynamics at hourly or sub-hourly resolution. This typically limits studies to between one and thirty climate years. This constraint creates an unavoidable need for subsampling \cite{pfenningerDealingMultipleDecades2017}: selecting a small number of climate years to represent the full distribution of meteorological conditions from the full reference dataset as well as possible \cite{nikMakingEnergySimulation2016}. \par

Several approaches to this climate year selection problem have emerged. One common strategy constructs synthetic "typical meteorological years" by making a composite of months or other intra-annual periods from different historical years \cite{ebrahimpourMethodGenerationTypical2010, kambezidisGenerationTypicalMeteorological2020}. Each representative period is selected to represent typical conditions for that period. While computationally convenient, such composite years introduce artificial discontinuities that disrupt the temporal coherence essential for modelling storage and hydropower reservoir dynamics \cite{hilbersImportanceSubsamplingImproving2019}. Moreover, there is considerable inter-annual variability in the meteorological parameters affecting the energy system \cite{pfenningerDealingMultipleDecades2017, vanderwielMeteorologicalConditionsLeading2019, grochowiczIntersectingNearoptimalSpaces2023, gotskeDesigningSectorcoupledEuropean2024, obergImpactInterannualWeather2025}. Typical meteorological years, or even shorter (aggregated) 'representative periods' \cite{nahmmacherCarpeDiemNovel2016, teichgraeberClusteringMethodsFind2019, hilbersImportanceSubsamplingImproving2019, hoffmannReviewTimeSeries2020} are therefore of limited use to accurately capture climate variability in resource adequacy or grid expansion studies for systems containing high shares of renewables. \par

An alternative approach, which has received comparatively little attention in the literature \cite{pfenningerDealingMultipleDecades2017}, is to select a (small) number of complete climate years from a larger climate modelling experiment \cite{nikMakingEnergySimulation2016, rugglesPlanningReliableWind2024, pecoraQuantifyingImpactsWeather2025}. The magnitude of climatic inter-annual variability is generally underestimated. It is so large that one cannot assume that using multiple complete years guarantees representativeness. Hence the usage of large ensembles with hundreds to thousands of years of data in climate science. A poorly chosen subset may over- or under-represent certain climatic conditions. For instance, selecting years that are consistently wind-rich while missing prolonged low-wind episodes could bias the energy system outcomes in the same way a single arbitrarily chosen year would. \par

The selection of climate years must therefore be guided by an explicit objective: the chosen subset should approximate as well as possible the full ensemble's multivariate, multi-seasonal distribution across all energy-relevant meteorological variables. Compared to composite approaches that connect months or shorter periods from different sources, using complete years limits discontinuities to the year boundaries alone. At the same time, selecting from a large pool of candidate years still offers sufficient freedom to closely approximate the target climatology, while only requiring relatively little years.\par

Finding an optimal, representative selection of multiple complete climate years is a combinatorial problem. A famous example of a combinatorial problem is called the `travelling salesman problem' \cite{hoffmanTravelingSalesmanProblem2001}. In such a problem, there is 'combinatorial explosion'. Selecting X years out of N years, with N $>>$ X, can be done in $N \choose X$ ways, which rapidly grows into an extreme amount of possibilities. To illustrate: selecting 5 years from 180 target years yields ${180 \choose 5} \approx 1.5 \times 10^{9}$ possible combinations, while selecting 30 years results in ${180 \choose 30} \approx 1.3 \times 10^{34}$ combinations. Therefore, exhaustive search (brute-force) is computationally infeasible, so an optimisation algorithm is needed. \par

Current practice of climate year selection in the energy sector varies considerably. For lack of validated alternatives, some studies employ random selection \cite{rugglesPlanningReliableWind2024, pecoraQuantifyingImpactsWeather2025} or pseudo-random selection, choosing years surrounding a target period without systematic optimisation \cite{EuropeanResourceAdequacy}. More sophisticated approaches include K-Medoids clustering \cite{TYNDP2024Scenarios, chatzistylianosAssessingImpactClimate2026}, clustering many climate years into a small amount of groups (clusters) and selecting the most representative member of each cluster. Finally, filtered random search methods are used, that evaluate a large number of random samples against selection criteria \cite{thoreySamplingRepresentativeYears}. Each approach involves trade-offs between computational cost, solution quality, and the ability to capture both typical conditions and extreme events. Some methods aim for criterion satisfaction with a minimal quality across all variables \cite{thoreySamplingRepresentativeYears}, while other studies aim for explicit optimisation \cite{hoffmannReviewTimeSeries2020}. \par

The aim of this study is to develop a method to robustly subsample a representative set of $X$ complete climate years from a larger reference dataset with $N$ climate years. The subsample should be as representative as possible of the reference dataset, in terms of the energy variables impacted by the weather. This representative set is a representative and validated input to detailed energy system models, given the constraint in computational capacity in those models. Therefore, using these representative climate years will result in an as-good-as-possible incorporation of climate variability into energy system modelling. \par

To determine the best-performing approach for upstream climate year selection, we review the previously mentioned existing climate year selection methods (random search, filtered random search, K-Medoids) and introduce simulated annealing as a new approach in this context. Based on this review we propose simulated annealing as the optimal approach (among those tested) for this application. Simulated annealing, a stochastic optimisation algorithm widely used for combinatorial problems \cite{vanlaarhovenSimulatedAnnealingTheory1987, vandorlandKNMINationalClimate}, offers an efficient means of exploring a large solution space while avoiding being trapped in local minima. We evaluate all methods on representativeness using the sliced Wasserstein distance \cite{villaniWassersteinDistances2009, frogner_learning_2015, kolouri_generalized_2019, condeixaWassersteinDistanceBasedTemporalClustering2020, solomon_wasserstein_nodate}, a metric from optimal transport theory that quantifies how well a subset represents the full (joint) multivariate distribution of a reference dataset. \par

Simulated annealing as a suitable method for representative climate year selection is agnostic to the application domain. Therefore, it would function equally well for selecting e.g. representative years for wind energy yield assessments \cite{pusatNewReferenceWind2021}, or for representative years for the built environment \cite{nikMakingEnergySimulation2016} or hydrological applications \cite{pechlivanidisInformationTheoryApproach2018}.


\section{Data}
\label{sec:data}
This study uses the Pan-European Climate Database (PECD), developed by the Copernicus Climate Change Service (C3S) in collaboration with ENTSO-E \cite{PECD2024, dubusFutureproofClimateDatabase2022}. The PECD dataset provides both meteorological variables from the climate simulations and their derived energy-impact. The climate simulations come from a selection of CMIP6 global climate models that have been statistically downscaled and bias-adjusted \cite{koivistoDevelopingSupportService2023, nayakValidationEuropeanWind2025}. The full projection database comprises simulations from 6 global climate models across 4 SSP scenarios, each spanning 85 years (2015-2100). \par

For our reference dataset, we select all 6 models under the SSP2-4.5 scenario for the period 2016-2045, yielding 180 climate years (6 models $\times$ 30 years). This 180-year dataset represents the full distribution of climate conditions from which we aim to select representative subsets. This means that our reference climate is centered around 2030, in a middle-of-the-road emission scenario. We assume there is no relevant forced climate signal or trend in the dataset. All variables are provided at hourly resolution and aggregated to NUTS0 level (national averages) \cite{RegionsEuropeanUnion2024}. However, to reduce computational burden in optimisation, we re-aggregate the data to daily means after all previous processing. For eventual use of the identified climate years in energy system models, users can use the native hourly resolution. \par

Because the energy impact of weather variables is non-linear, we perform climate year selection on the derived energy variables rather than raw meteorological fields. For example, wind turbine output scales with the cube of wind speed \cite{sohoniCriticalReviewWind2016, nayakValidationEuropeanWind2025}. Selecting years based on energy-impact variables ensures representativeness on the variables that matter most for energy system modelling. \par

The PECD contains capacity factors for wind and solar generation, which take these non-linear relations into account. We convert the capacity factors to energy production values using installed capacities based on a 2030 European scenario from the European Resource Adequacy Assessment 2024 by ENTSO-E \cite{EuropeanResourceAdequacy}. Electricity demand exhibits a U-shaped dependence on temperature due to heating and cooling loads \cite{bessecNonlinearLinkElectricity2008}. We use the population-weighted temperature from the PECD and a simple demand-temperature relationship to calculate daily national electricity demand \cite{vandermostExtremeEventsEuropean2022}. We take the hydropower production values (run-of-river with and without pondage and reservoir hydropower) at face value, assuming no changes in capacity before 2030. The resulting variables used for climate year selection are: offshore wind production, onshore wind production, solar production, (pondage) run-of-river hydropower production, reservoir hydropower production, electricity demand, and residual load (defined as demand minus total renewable production). The usage of load/generation values in GWh/day ensures that countries with large load or generation are weighted more heavily in calculating the representativeness. \par 

Importantly, the algorithms work identically on other datasets, whether they consist of climate variables (e.g. wind speeds, solar radiation, temperature), capacity factors, or generation (or other) impact variables as used here. 


\section{Method}
\label{sec:methods}
To demonstrate and compare subsampling methods for representative climate year selection, we performed three case studies, creating two target subset sizes representative over two different spatial domains. A larger subsample of 30 climate years was selected to support resource adequacy studies such as ENTSO-E's European Resource Adequacy Assessment (ERAA). A smaller subsample of 5 climate years was selected for computationally intensive applications such as investment planning studies like ENTSO-E's Ten Year Network Development Plan (TYNDP) or national investment studies, where fewer climate years can be processed. We evaluated four different subsampling methods, described in Section \ref{subsec:methods_selection_algorithms}. The representativeness of the subset was quantified using a cost function (Section~\ref{subsec:methods_costfunction}), which assesses the joint, multi-variate distribution of weather-impacted energy generation/demand variables.

\subsection{Cost function}
\label{subsec:methods_costfunction}
To evaluate the representativeness of a subsample, we used the sliced Wasserstein distance (SWD) as our cost function. The Wasserstein distance is a concept from optimal transport theory \cite{villaniWassersteinDistances2009} and is frequently used in data science and machine learning applications \cite{kolouri_generalized_2019, frogner_learning_2015, solomon_wasserstein_nodate}. It quantifies the minimum ``work'' required to transform one probability distribution into another. It is also more intuitively known as the Earth mover's distance. Crucially, the Wasserstein distance accounts for the joint dependence structure between variables, making it well-suited for evaluating the multivariate climate and energy data of interest here, where correlations between the variables are important. In the Supplementary Information Section~\ref{subsec:SI_SWD_illustrations}, we provide an elaborate justification of the usage of the SWD as the adequate metric to judge representativeness, by comparing it to other common scoring metrics via a set of illustrative examples. \par

As described by \textcite{francoisMultivariateBiasCorrections2020}, the Wasserstein distance measures the distance between two multivariate probability distributions. Here that means subset distribution $P$ consisting of $X$ climate years and reference set distribution $Q$ consisting of all $N$ years. The Wasserstein distance is the minimum cost of transforming subset $P$ into reference set $Q$. The metric has previously been applied in climate and energy contexts \cite{vissio_evaluating_2020, hahmannMakingNewEuropean2020}. For example, \textcite{hahmannMakingNewEuropean2020} used the (one-dimensional) Earth mover's distance to study the similarity between simulated and observed wind speeds and directions. \par

Computing the exact Wasserstein distance becomes computationally expensive for high-dimensional data and large data sets (180 years, 8 variables, daily values). We therefore used the sliced Wasserstein distance (SWD) as a computationally efficient approximation \cite{bonneelSlicedRadonWasserstein2015}. This approach projects the multivariate data onto $N_p$ random one-dimensional projections, computes the one-dimensional Wasserstein distance for each projection, averages the squared results and takes the square root. The SWD has previously been applied in climate settings, e.g. to quantify uncertainty in climate projections \cite{harrisQuantifyingUncertaintyClimate2024}. \par

Following the mathematical formulation of \textcite{harrisQuantifyingUncertaintyClimate2024}, the SWD between $n$-dimensional distributions $P$ and $Q$ is defined as:
\begin{equation}
    \text{SWD}(P,Q) = \left( \int_{\mathbb{S}^{n-1}} W_2^2\bigl(\mathcal{R}_\theta[P], \mathcal{R}_\theta[Q]\bigr)\, d\theta \right)^{1/2}
\label{eq:SWD}
\end{equation}
where $\theta \in \mathbb{S}^{n-1}$ is a unit direction, $\mathcal{R}_\theta[P]$ and $\mathcal{R}_\theta[Q]$ are the one-dimensional projections of $P$ and $Q$ along $\theta$, and $W_2$ is the two-Wasserstein distance. For the projected one-dimensional distributions $F = \mathcal{R}_\theta[P]$ and $G = \mathcal{R}_\theta[Q]$, the two-Wasserstein distance is given by:
\begin{equation}
    W_2(F,G) = \bigl\| F^{-1} - G^{-1} \bigr\|_2
\end{equation}
where $F^{-1}$ and $G^{-1}$ are the quantile functions. For this study, the sliced Wasserstein distance was computed using the Python Optimal Transport library \cite{flamary2024pot}. \par
Finally, the seasonal timing of climate conditions matters for energy system applications: a low-wind period in summer has different system implications than one in winter, particularly for resource adequacy assessments. A metric based on the annual distribution alone could mask seasonal biases, for example by allowing an overly windy summer to compensate for little wind in winter. To prevent such compensation across seasons, we introduce the seasonal sliced Wasserstein distance (SSWD). Each dataset was partitioned into an extended winter (October-March) and summer (April-September) period, and the SWD was computed separately for each season. The SSWD was then defined as the maximum of the two seasonal distances:
\begin{equation}
    \text{SSWD}(P, Q) = \max\bigl(\text{SWD}(P_{\text{winter}}, Q_{\text{winter}}),\; \text{SWD}(P_{\text{summer}}, Q_{\text{summer}})\bigr)
\label{eq:SSWD}
\end{equation}
where $P_{winter}$ and $Q_{winter}$ denote the winter portions of the subset and reference distributions, and $P_{summer}$ and $Q_{summer}$ their summer counterparts. By optimising to minimise the worst-performing season, the SSWD ensures that the selected subset adequately represents both seasonal distributions rather than achieving a low overall cost at the expense of one season. This effect is illustrated in Supplementary Information Section~\ref{subsec:SI_SWD_illustrations}. The unit of the (S)SWD follows that of the underlying distributions, so in this study it is in GWh/day. \par

Due to the random nature of the projections, the (S)SWD carries some uncertainty that decreases with increasing $N_p$ (S.I. Figure~\ref{fig:S5_sswd_uncertainty_analysis_NL30y}-\ref{fig:S7_sswd_uncertainty_analysis_EU5y}). We used a two-stage approach to balance computational efficiency and accuracy. During the initial search phase, we used $N_p = 50$ projections to enable rapid evaluation of many candidate subsamples. This gave a coefficient of variation in the order of 5~\%.  For final calculations and for subsamples identified as promising (best 10~\%) based on the initial search, we used $N_p = 200$ projections to obtain more accurate distance estimates. This gave a coefficient of variation of about 2--3~\%. \par

In summary, the subset of climate years with the lowest seasonal sliced Wasserstein distance with respect to the full reference dataset was considered to be the most representative.

\subsection{Selection algorithms}
\label{subsec:methods_selection_algorithms}
In this subsection, we briefly introduce the different compared selection algorithms. The algorithms are further detailed in pseudo-code in Supplementary Information Section \ref{subsec:SI_selection_algorithms}.

\subsubsection{Random search}
\label{subsec:methods_random_draws}
The simplest approach to climate year selection is random search, which here serves as a baseline for comparison with more sophisticated methods. In random search, we simply sampled $X$ complete years at random without replacement from the reference dataset, and then calculated the cost function value. Despite, or maybe because of, its simplicity, random or near-random selection remains common in practice \cite{EuropeanResourceAdequacy, TYNDP2024Scenarios, rugglesPlanningReliableWind2024}. For example, selecting a fixed number of years centred around a target year effectively constitutes a random draw, since the weather conditions in those years are stochastically determined by internal climate variability. However, note that the random search algorithm implemented here consists of many random draws, scoring them, and keeping the best in terms of the SSWD (see Algorithm~\ref{alg:random_draw}). This is much more robust than simply taking one random draw, using that, and hoping for good luck, as is often still done in practice. \par

\subsubsection{Filtered random search}
\label{subsec:methods_frs}
Filtered random search (FRS) extends the random search approach (Section~\ref{subsec:methods_random_draws}) by introducing a sequential pre-selection step that filters candidate subsets based on cheap-to-calculate scores before computing the full cost function. The approach used here (see Algorithm~\ref{alg:frs}) was adapted from the approach by \textcite{thoreySamplingRepresentativeYears}. \par

For each candidate subset $\mathcal{S}^{(m)}$, cheap scores were evaluated sequentially per variable $v$ and per season $t$. First the mean difference $d_{\mathrm{mean}}(\mathcal{S}^{(m)}, \mathcal{D}, v, t)$, and then the energy distance score $d_{\mathrm{energy}}(\mathcal{S}^{(m)}, \mathcal{D}, v, t)$ were computed. The mean difference is the absolute difference between the subset and reference means per variable, per season. The energy distance is a statistical distance between the subset and the reference that measures univariate distributional similarity beyond the mean, also accounting for spread and shape \cite{szekelyEnergyStatisticsClass2013}. Both scores are computed per variable and per season, and are substantially cheaper to evaluate than the SSWD, which makes them suitable as pre-filters. After each score was computed for a variable and season, it was immediately compared against a per-variable, per-season threshold. If the threshold was exceeded, evaluation of that candidate subset was abandoned and the next candidate was drawn. \par

Only if all scores for all variables passed their respective thresholds was the true cost $\mathrm{SSWD}(\mathcal{S}^{(m)}, \mathcal{D})$ computed. This hierarchical and early-stopping evaluation allows a much larger number of candidate subsets to be assessed within the same computational budget, as the expensive SSWD calculation is reserved for the most promising candidates. \par

Prior to the main FRS loop, a threshold-tuning step was performed to calibrate the per-variable, per-season thresholds for $d_{\mathrm{mean}}$ and $d_{\mathrm{energy}}$. Thresholds were set such that approximately one in every thousand randomly drawn subsets passes the full pre-selection, balancing the strictness of filtering against the number of subsets that reach the SSWD evaluation. \par

\subsubsection{K-Medoids}
\label{subsec:methods_k_medoids}

K-Medoids clustering is currently used by ENTSO-E for climate year selection in the Ten-Year Network Development Plan (TYNDP) 2024 \cite{TYNDP2024Scenarios} and has been applied in some form in various energy system studies \cite{teichgraeberClusteringMethodsFind2019, hoffmannReviewTimeSeries2020, chatzistylianosAssessingImpactClimate2026}. The method is closely related to K-means clustering, with one key difference. K-means defines cluster centres as the mean location of all points in a cluster, which does not necessarily correspond to an actual data point. K-Medoids, on the other hand, selects an actual data point from within each cluster as its representative. This property makes K-Medoids particularly suitable for climate year selection, as each cluster is represented by a real climate year rather than an artificial composite. \par

The features used for K-Medoids clustering are the seasonal means and standard deviations of all energy-impact variables on the country level. To select $X$ representative climate years, we calculated $X$ clusters and therefore also $X$ medoids representing that cluster. Those $X$ medoid years combined constitute the identified subset. For each run, we repeated this K-Medoids clustering many times with stochastic K-Medoids++ initialisation, and kept the best, median and worst subset in terms of the SSWD (see Algorithm~\ref{alg:k_medoids}) per run.

\subsubsection{Simulated annealing}
\label{subsec:methods_simulated_annealing}
Simulated annealing (SA) is a stochastic optimisation algorithm that tries to find the global minimum of a cost function over a large discrete solution space  \cite{kirkpatrickOptimizationSimulatedAnnealing1983, vanlaarhovenSimulatedAnnealingTheory1987}. Unlike K-Medoids, which partitions the solution space into clusters, SA directly optimises for representativeness by iteratively improving a candidate subset. It is well-suited for combinatorial problems such as climate year selection \cite{vandorlandKNMINationalClimate}, and to our knowledge has not previously been applied in this context. \par

The algorithm starts from a randomly drawn subset $\mathcal{S}$ as its initial solution, and iteratively perturbs it by swapping one or more of its years. Improvements in terms of cost are accepted unconditionally, and worse solutions are accepted with a probability that decreases as the temperature of the algorithm $T$ declines. Here, `temperature' is a metaphorically named hyperparameter of SA that governs the algorithm's willingness to accept worse solutions. Its value is unrelated to the climate temperatures in the underlying data. The cooling mechanism of SA drives convergence, while the probabilistic acceptance prevents the algorithm from being trapped in local minima. A reheating mechanism provides additional protection against premature convergence, if the algorithm has not found an improved solution for $n_{stall}$ iterations. The full procedure is detailed in Algorithm~\ref{alg:sa}.

\subsection{Year definition}
\label{subsec:methods_year_definition}
Climate years are defined as hydrological years, running from 1 April to 31 March. For example, the hydrological year 2020 runs from 1 April 2020 to 31 March 2021. This convention preserves the integrity of the winter season. This is important for resource adequacy assessments, as winter generally represents the period of highest system stress. Splitting the winter across two calendar years might break up resource adequacy stress events, and would introduce artificial discontinuities in hydropower reservoir levels. On April 1st, water reservoirs are usually at or near their annual minimum before they start to fill in spring \cite{wolfgangHydroReservoirHandling2009}, so this is a natural reset point where inter-annual variation is relatively small.

\subsection{Experiment design}
\label{subsec:methods_experiment_design}
We performed three case studies and compared each method across them. We began by selecting a representative subset of $X = 30$ years for a single country (the Netherlands), targeting resource adequacy studies. Next, we used the same subset size but extended the domain to all countries in Europe. This is the most complex case and directly relevant for continental-scale resource adequacy studies such as ERAA. We compared our 30-year selections for the four methods, and with ERAA25 \cite{EuropeanResourceAdequacy} as a `current practice' reference. Finally, we used the same European domain and selected $X = 5$ representative years, aiming for a representative set for European investment planning type studies. Together, these three experiments allowed us to evaluate method performance across both small and large spatial scales and across different subset sizes. \par

For each case study, every method was run 25~times to reduce the influence of stochastic initialisation. We recorded the best, median, and worst SSWD score across runs for each method. To quantify consistency, we defined two complementary metrics. The \textit{luck factor} ($\mathcal{L}$) measures how much better the best run of a method is relative to the median run of random search. We used random search as a common normalisation baseline, so that the luck factor can be compared between methods. The \textit{luck factor} is defined as:
\begin{equation}
    \mathcal{L} = \frac{\tilde{c} - c^*}{\tilde{c}_{rand}}
    \label{eq:luck_factor}
\end{equation}
and the \textit{fragility factor} ($\mathcal{F}$) measures how much worse the worst run of that method is relative to the median run of random search:
\begin{equation}
    \mathcal{F} = \frac{c_{\max} - \tilde{c}}{\tilde{c}_{rand}}
    \label{eq:fragility_factor}
\end{equation}
where $c^*$ is the best SSWD score, $\tilde{c}$ the median score, and $c_{\max}$ the worst score across the 25 runs for each method. $\tilde{c}_{rand}$ is the median score of the random search method. A method with low $\mathcal{L}$ and low $\mathcal{F}$ is consistent regardless of initialisation, while high values indicate that outcomes depend heavily on initial conditions or luck. \par

Simulated annealing was run with exponential cooling at rate $c_r = 0.9975$, with reheating by a factor of $1.8$ after $n_{\mathrm{stall}} = 300$ iterations without improvement. We used $T_{start} = 2$, and $T_{end} = 10^{-3}$. This gave a minimum of 
\begin{equation*}
    n_{min} = \frac{\ln(T_{end}/T_{start})}{\ln(c_r)} = \frac{\ln(10^{-3}/2)}{\ln(0.9975)} = 3037
\end{equation*}
iterations, but in practice this was often higher due to reheating at stalls. We used a maximum of $it_{\max} = 5000$ iterations per SA run.

K-Medoids and random search were each run for the same number of iterations as simulated annealing used, rounded up to the nearest multiple of 25 due to 25 repeated runs per method. This ensured that the only differences between compared methods were algorithmic: SA uses local search, K-Medoids identifies medoid years within clusters, and random search samples independently each iteration. Filtered random search was allowed ten times as many iterations as the other methods. Its hierarchical pre-selection on cheap metrics ensured that only a small fraction of candidates (order 1-in-1000) reached the expensive SSWD evaluation. This made the total computational cost roughly equivalent across all methods.

\section{Results}
\label{sec:results}

\subsection{Case 1: Single country, 30 representative years}
\label{subsec:results_NL30y}
In this first case, we aim to select a subset of $X = 30$ representative years for a single country, the Netherlands. Representativeness is assessed on 1 country $\times$ 6 variables. Note that there are 8 relevant energy variables in PECD, but there is no pondage run-of-river and reservoir hydropower in the Netherlands. The experiment follows the design outlined in Section \ref{subsec:methods_experiment_design}. \par

\begin{figure}[!t]
    \centering
    \includegraphics[width=0.95\linewidth]{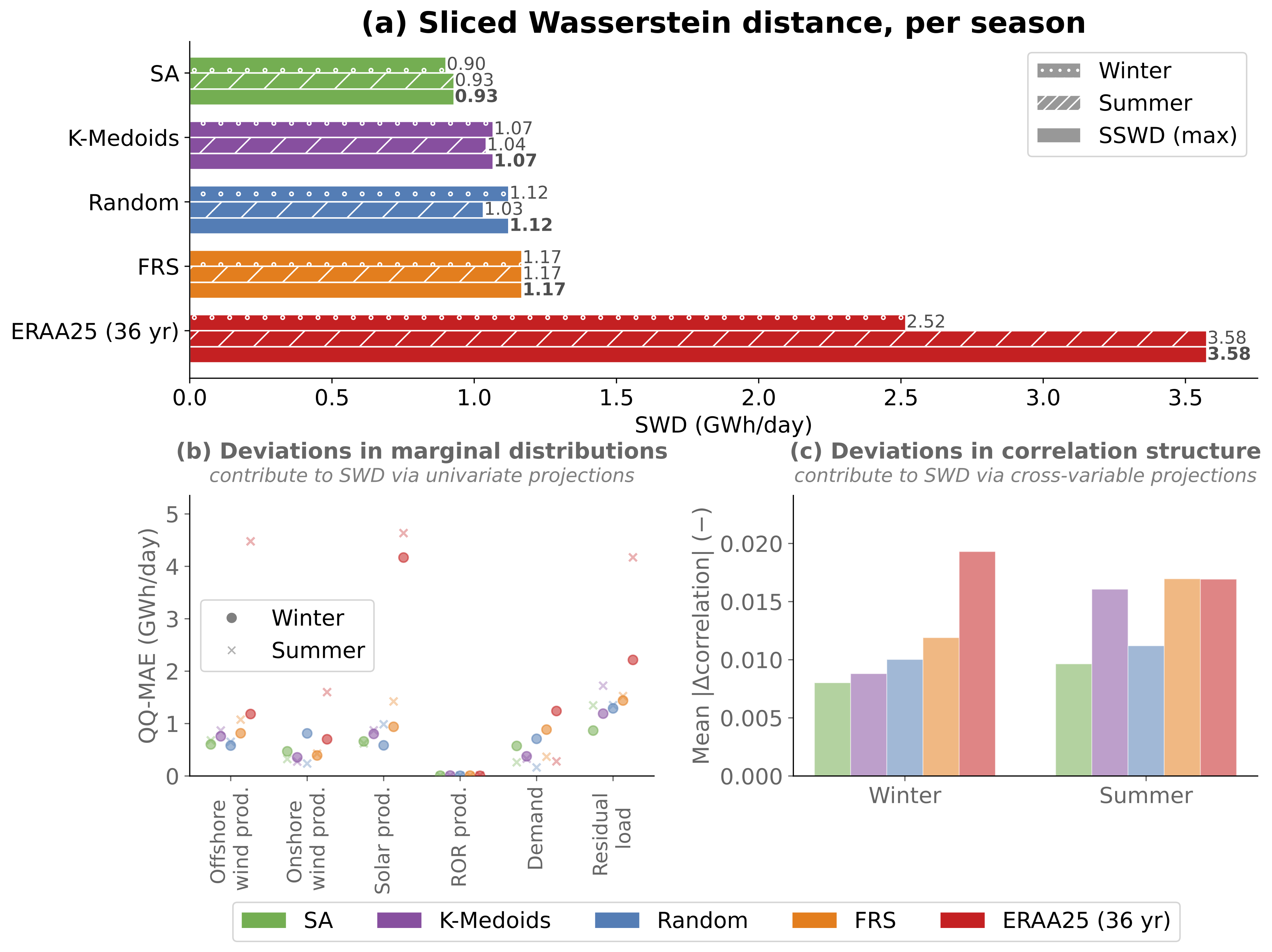}
    \caption{Representativeness of a subset of $X = 30$ selected climate years for the Netherlands across four selection methods and the ERAA25 benchmark (36 years). \textbf{(a)} Sliced Wasserstein Distance (SWD) per season, decomposed into extended winter (October-March) and summer (April-September). Bold values indicate the SSWD (seasonal maximum). \textbf{(b)} Deviations in marginal distributions per variable, expressed as the quantile-quantile Mean Absolute Error (GWh/day). \textbf{(c)} Deviations in inter-variable correlation structure with respect to the reference. In all panels, lower values indicate greater representativeness. Panel (a) represents the main representativeness score, panel (b) and (c) are components of this score.}
    \label{fig:fig1_case1_summary}
\end{figure}

Simulated annealing (SA) yields the most representative subset of all four methods (Figure~\ref{fig:fig1_case1_summary}a). The second-best method (K-Medoids) has a 15~\% higher best score across its runs, and the worst performing method (filtered random search) has a~26~\% higher score. It is striking that simulated annealing achieves a more than 3.5 times lower SSWD score than the ERAA25 subset currently used for the European Resource Adequacy Assessment \cite{EuropeanResourceAdequacy}, despite ERAA25 being at advantage by containing 36~years rather than 30. \par

The superiority of SA holds across both components of the SSWD: accuracy of marginal distributions (Figure~\ref{fig:fig1_case1_summary}b) and inter-variable correlation structure (Figure~\ref{fig:fig1_case1_summary}c). This confirms that SA produces a comprehensively more representative subset than the other tested methods, not merely one that scores well on a single aspect of representativeness.

\begin{figure}[!t]
    \centering
    \includegraphics[width=0.85\linewidth]{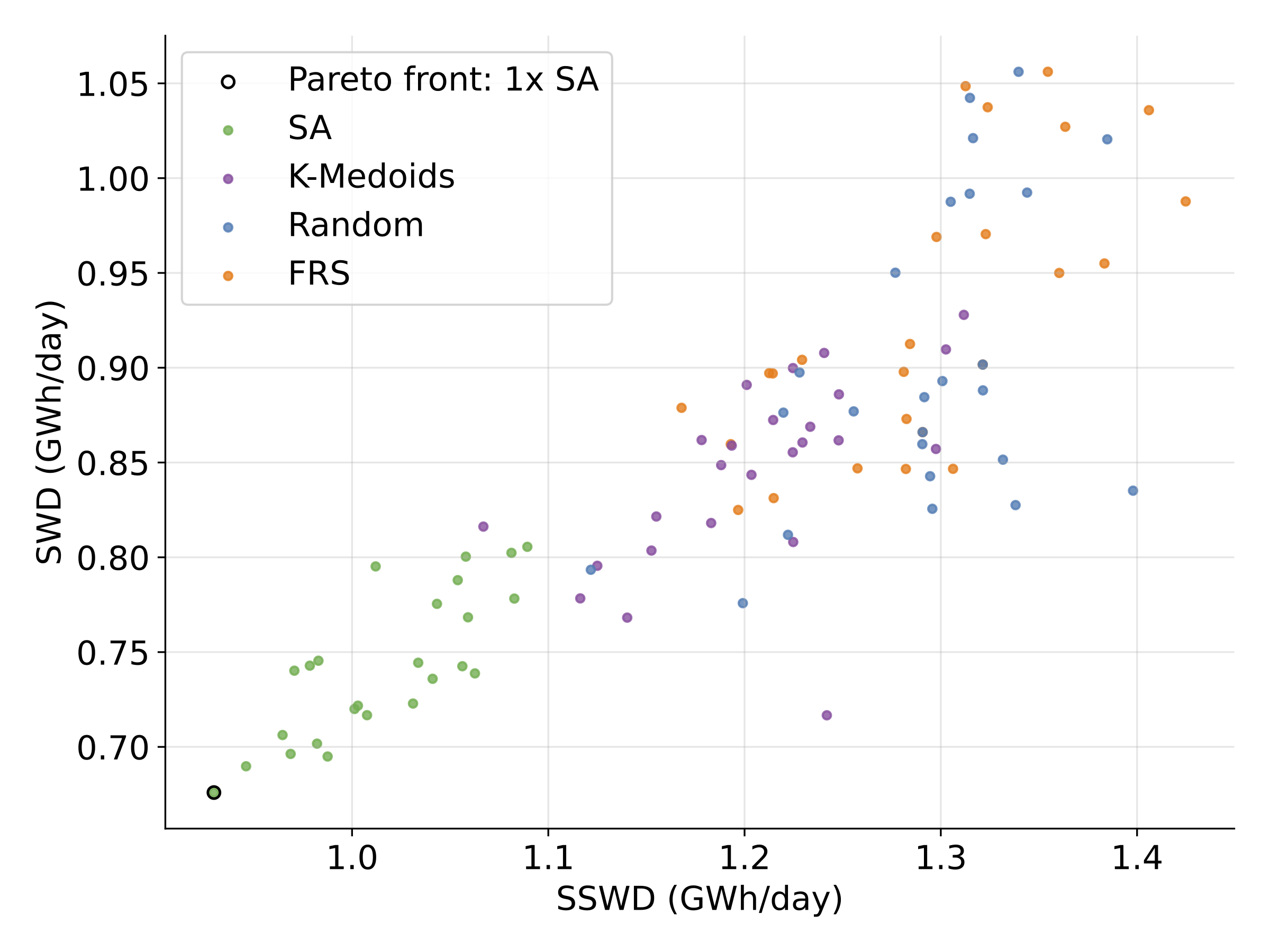}
    \caption{SSWD (seasonal) versus SWD (annual) scores across all 25 runs per method, for $X = 30$ years selected for the Netherlands. Each dot represents one run. The Pareto front, solutions where neither metric can be improved without degrading the other, consists of a single SA run, which is therefore the best-scoring subset overall.}
    \label{fig:fig2_case1_pareto}
\end{figure}

SA dominates across both seasonal and annual representativeness metrics, and does so consistently across runs (Figure~\ref{fig:fig2_case1_pareto} and Table \ref{tab:tab1_case1_summary_table}). Note that annual scores are shown for reference, but do not form the main optimisation objective. SA runs cluster in the lower-left corner of the $SSWD-SWD$ space, indicating both high solution quality and low sensitivity to initialisation. The Pareto front, solutions where neither metric can be improved without degrading the other, consists of a single SA run, the best-scoring subset. The other methods show higher scores and higher luck and fragility factors, indicating both worse solutions and higher dependence on luck to get good scores. Of the 100 candidate subsets (25 runs $\times$ 4 methods), the top 22 all come from SA.

The best selected subset achieves an SSWD equivalent to a median-quality random draw of 140 years. This means that the SA climate year selection achieves an 'effective sample size' that is $4.67 \times$ its actual size. \par

\begin{table}[!hb]
    \centering
    \small
    \setlength{\tabcolsep}{4pt}
    \caption{Case 1: Summary table with performance and reliability per method. SSWD scores in GWh/day. SA = simulated annealing, RS = random search, FRS = filtered random search. Median best-of-run: median of the best SSWD score found in each of the 25 runs (i.e.\ $\tilde{c}$ in Eqs.\ \ref{eq:luck_factor}--\ref{eq:fragility_factor}). Median single-draw: median SSWD across all candidates evaluated within a run, reported as the min--max range over the 25 runs; it represents the typical score a user would obtain from a single iteration of the method. SA has no meaningful single-draw value, as its trajectory moves from random initialisation to converged solution.}
    \label{tab:tab1_case1_summary_table}
    \begin{tabular}{lrrrr}
    \toprule
     & \textbf{SA} & \textbf{K-Medoids} & \textbf{RS} & \textbf{FRS} \\
    \midrule
    Best overall score (SSWD)     & \textbf{0.93}    & 1.07       & 1.12       & 1.17        \\
    Rank of best score (of 100)   & \textbf{1}       & 23         & 28         & 33          \\
    Median best-of-run            & \textbf{1.01}    & 1.21       & 1.30       & 1.29        \\
    Median single-draw            & N/A              & 2.42--2.47 & 2.89--2.97 & 1.60--1.73  \\
    Luck factor                   & \textbf{0.06}    & 0.11       & 0.14       & 0.09        \\
    Fragility factor              & \textbf{0.06}    & 0.07       & 0.07       & 0.10        \\
    \midrule
    Total iterations              & 120{,}082        & 120{,}100  & 120{,}100  & 1{,}200{,}825 \\
    Number of runs                & 25               & 25         & 25         & 25          \\
    Iterations per run            & 3{,}742--5{,}000 & 4{,}804    & 4{,}804    & 48{,}033    \\
    \bottomrule
    \end{tabular}
\end{table}

\subsection{Case 2: Whole Europe, 30 representative years}
\label{subsec:results_Europe30y}

For the second case, we extend the domain to all countries in the PECD domain, which contains the majority of Europe. Here, we again select a subset of $X = 30$ representative years, the same amount of years as in Case 1 (Section~\ref{subsec:results_NL30y}). Here, we check if the sampling methods also work for a larger spatial domain and can find representative datasets in a significantly higher dimensionality. We now aim for representativeness across 8 variables $\times$ 35 countries = 280 country-level variables of which 211 are non-zero. The remaining 69 variables are discarded. They correspond to countries without installed capacity for a given renewable technology. \par

\begin{figure}[h]
    \centering
    \includegraphics[width=0.95\linewidth]{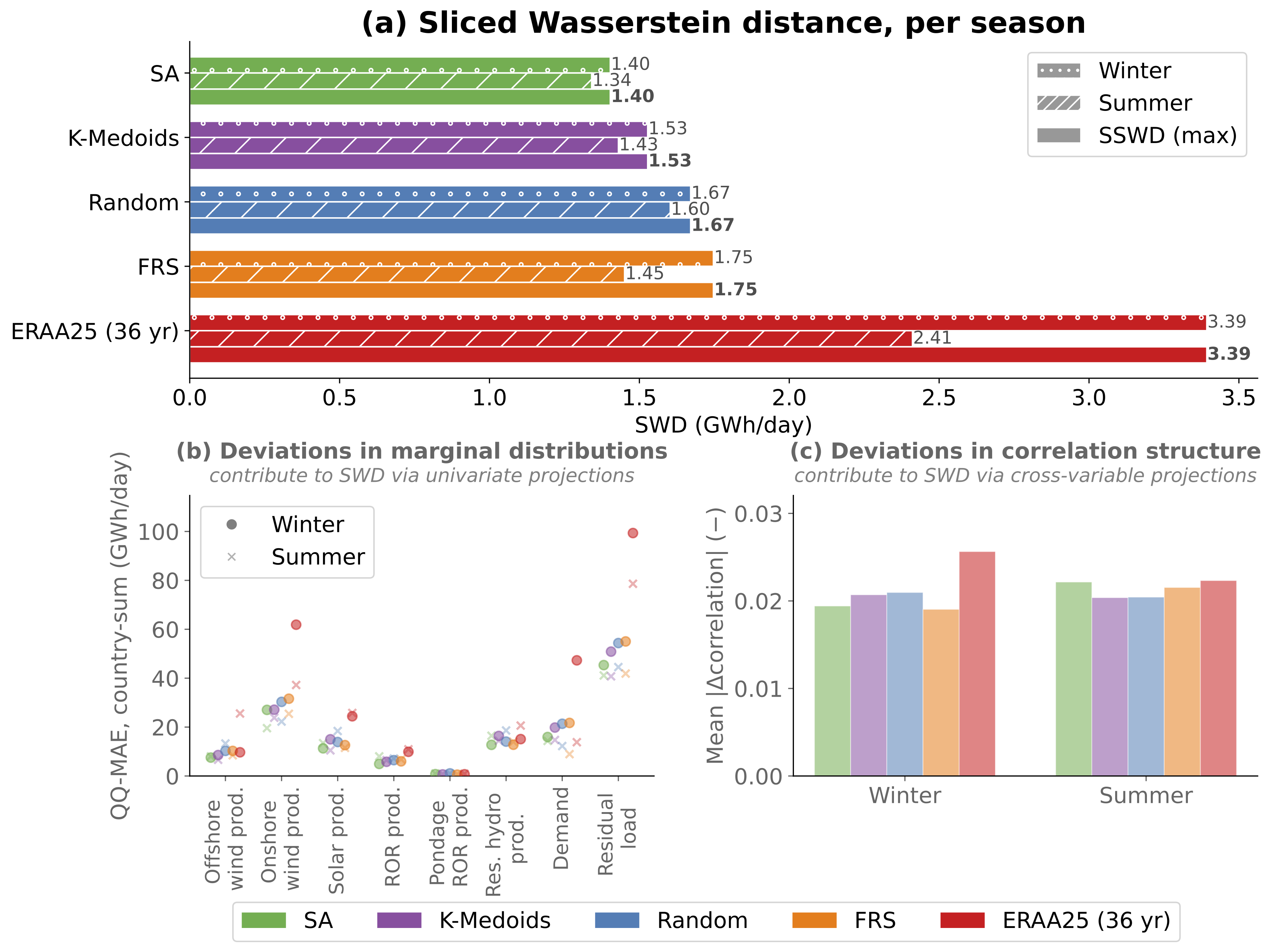}
    \caption{As Figure~\ref{fig:fig1_case1_summary}, but here showing the representativeness of $X = 30$ selected climate years for Europe across four selection methods and the ERAA25 benchmark (36 years).}
    \label{fig:fig3_case2_summary}
\end{figure}

Simulated annealing again yields the most representative subset of all four methods (Figure~\ref{fig:fig3_case2_summary}a). The second best method (K-Medoids) has a 9~\% higher best cost across its 25 runs, and the worst performing method (filtered random search) has a 25~\% higher best cost. For this case, the pre-filtering step in FRS worsens results compared to random search. With many countries and variables, even a good subset with low overall SSWD may exceed a single threshold for one variable in one country and be discarded. The filtering therefore does more harm than good for SSWD optimisation in cases with many countries. \par

SA dominates across both seasonal and annual representativeness metrics, and does so consistently across runs (Table~\ref{tab:tab2_case2_summary_table} and S.I. Figure~\ref{fig:figS7_case2_pareto}). Of the 100 candidate subsets, the top 18 results are all simulated annealing. Its median score represents a larger improvement over the other methods than its best score, indicating low dependence on initialisation luck. Random search also shows a low luck factor, but consistently achieves higher cost function values.

\begin{table}[hb]
\centering
\small
\setlength{\tabcolsep}{4pt}
\caption{As Table~\ref{tab:tab1_case1_summary_table}, but for Case 2.}
\label{tab:tab2_case2_summary_table}
\begin{tabular}{lrrrr}
\toprule
 & \textbf{SA} & \textbf{K-Medoids} & \textbf{RS} & \textbf{FRS} \\
\midrule
Best overall score (SSWD)     & \textbf{1.40}    & 1.53       & 1.67       & 1.75          \\
Rank of best score (of 100)   & \textbf{1}       & 19         & 39         & 63            \\
Median best-of-run            & \textbf{1.48}    & 1.66       & 1.74       & 1.89          \\
Median single-draw            & N/A              & 2.99--3.03 & 3.52--3.60    & 1.75\textsuperscript{*}--3.33    \\
Luck factor                   & \textbf{0.04}    & 0.08       & \textbf{0.04} & 0.08       \\
Fragility factor              & \textbf{0.05}    & 0.06       & 0.10       & 0.82          \\
\midrule
Total iterations              & 123{,}657        & 123{,}675  & 123{,}675  & 1{,}236{,}575 \\
Number of runs                & 25               & 25         & 25         & 25            \\
Iterations per run            & 4{,}211--5{,}000 & 4{,}947    & 4{,}947    & 49{,}470      \\
\bottomrule
\end{tabular}
\\ \flushleft \textsuperscript{*}For this run, only one candidate passed the pre-filter, and this candidate by chance gives the best overall score.
\end{table}

The best selected subset achieves an SSWD equivalent to a median-quality random draw of 130 years. SA climate year selection thus achieves an 'effective sample size' that is $4.33$ times its actual size.

\clearpage
\subsection{Case 3: Whole Europe, 5 representative years}
\label{subsec:results_Europe5y}

For the final case, we select a smaller subset of $X = 5$ representative years for the same region as Case 2, the majority of Europe (whole PECD domain). Here we test if the methods are robust to selecting smaller representative subsets. \par

\begin{figure}[!hb]
    \centering
    \includegraphics[width=0.9\linewidth]{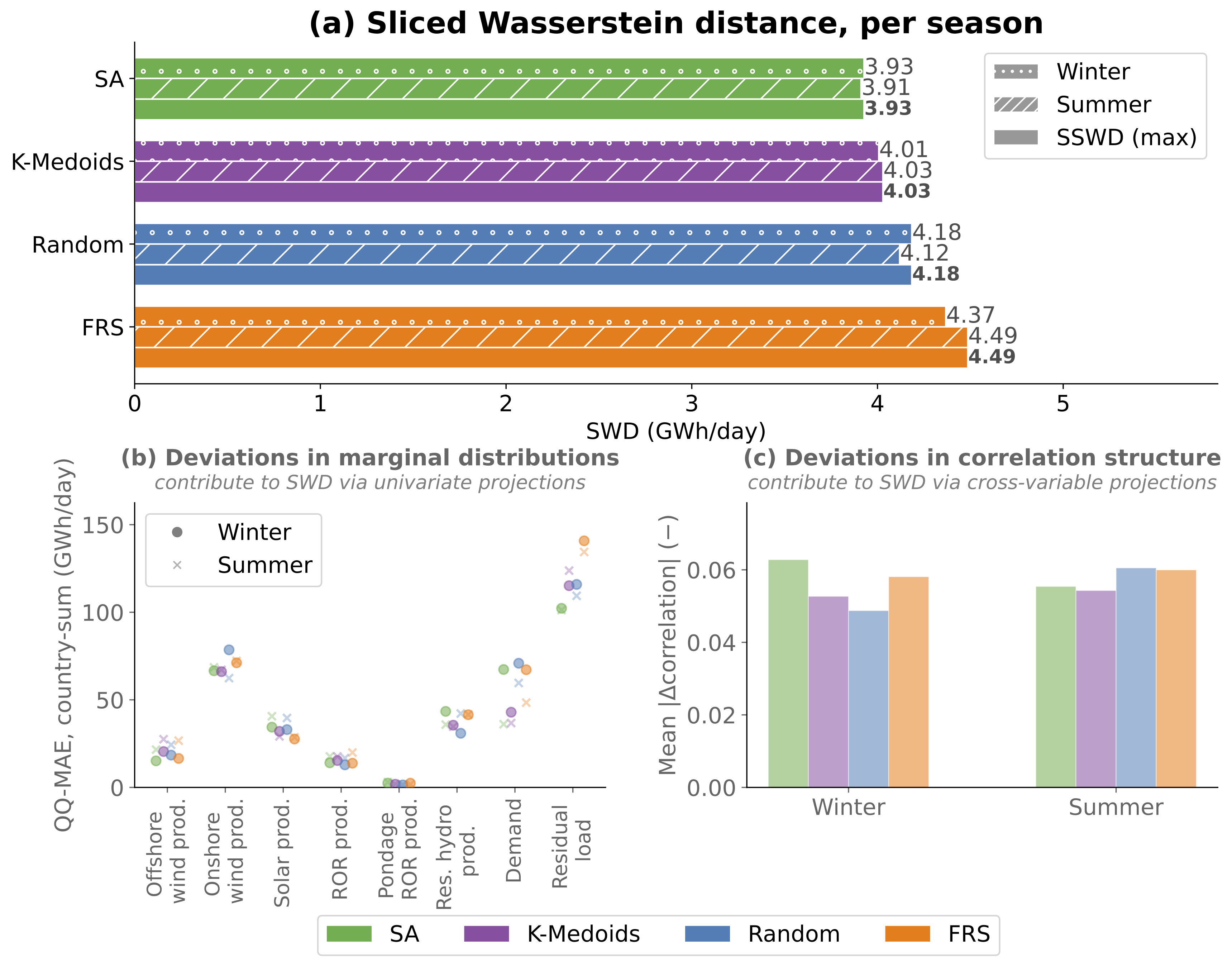}
    \caption{As Figure~\ref{fig:fig1_case1_summary}, but here showing the representativeness of $X = 5$ selected climate years for Europe across four selection methods. Note that ERAA25 \cite{EuropeanResourceAdequacy} does not provide a 5-year subset and hence is not included in the comparison.}
    \label{fig:fig5_case3_summary}
\end{figure}

Indeed, simulated annealing again produces the most representative subset (Figure \ref{fig:fig5_case3_summary}a), though the margin over other methods is smaller than in the 30-year cases (Figure~\ref{fig:fig1_case1_summary}a \& \ref{fig:fig3_case2_summary}a). The best selected subset achieves an SSWD equivalent to a median-quality random draw of roughly 25 years, an effective sample size that is approximately 5 times its actual size. K-Medoids achieves a best SSWD score only 3~\% higher than SA, compared to 9~\% in Case~2. Random search scores 6~\% higher and filtered random search 14~\% higher. The reduced gap is expected, with only 5 years to select from 180, the solution space is much smaller $({180 \choose 5} \approx 1.5 \times 10^{9}$, whereas ${180 \choose 30} \approx 1.3 \times 10^{34})$, giving less room for the algorithm to differentiate itself. The optimisation landscape is likely also flatter near the optimum. With only 5 years to represent all of Europe, many distinct subsets achieve similar SSWD scores, reducing the advantage of a more sophisticated search method. \par

Despite the smaller margins in best scores, SA remains the most consistent method (Table~\ref{tab:tab3_case3_summary_table} and S.I. Figure~\ref{fig:figS8_case3_pareto}). The top 3 of 100 candidate subsets are all from simulated annealing, and its median score is 4~\% better than K-Medoids and 8~\% better than random search. SA also achieves the lowest luck factor (0.07) and fragility factor (0.05) of all methods, indicating stable performance regardless of initialisation. Filtered random search, by contrast, shows the highest fragility (0.36) and the widest solution spread in supplementary Figure~\ref{fig:figS8_case3_pareto}. This confirms that pre-filtering continues to be detrimental at European scale, also for a smaller size representative set.

\begin{table}[ht]
\centering
\small
\setlength{\tabcolsep}{4pt}
\caption{As Table~\ref{tab:tab1_case1_summary_table}, but for Case 3.}
\label{tab:tab3_case3_summary_table}
\begin{tabular}{lrrrr}
\toprule
 & \textbf{SA} & \textbf{K-Medoids} & \textbf{RS} & \textbf{FRS} \\
\midrule
Best overall score (SSWD)     & \textbf{3.93}    & 4.03       & 4.18          & 4.49          \\
Rank of best score (of 100)   & \textbf{1}       & 4          & 13            & 54            \\
Median best-of-run            & \textbf{4.26}    & 4.39       & 4.52          & 5.05          \\
Median single-draw            & N/A              & 7.52--7.66 & 9.34--9.49    & 5.38--7.18    \\
Luck factor                   & \textbf{0.07}    & 0.08       & \textbf{0.07} & 0.13          \\
Fragility factor              & \textbf{0.05}    & 0.07       & 0.08          & 0.36          \\
\midrule
Total iterations              & 125{,}000        & 125{,}000  & 125{,}000     & 1{,}250{,}000 \\
Number of runs                & 25               & 25         & 25            & 25            \\
Iterations per run            & 5{,}000          & 5{,}000    & 5{,}000       & 50{,}000      \\
\bottomrule
\end{tabular}
\end{table}

\section{Discussion}
\label{sec:discussion}
A key implication of this study is that subset representativeness \textit{can} and should be objectively quantified and optimised. This stands in contrast to current practice in many energy system studies. Climate year subsets are typically selected without any explicit representativeness criterion: a fixed number of years is drawn around a target year, or a historical period is used out of convention. Such choices are effectively single random draws, not validated against the underlying climate distribution. The cost of doing so is substantial: ENTSO-E's ERAA25 subset \cite{EuropeanResourceAdequacy} scores almost 2.5 times worse on SSWD than the SA-optimised 30-year subset for Europe (Figure~\ref{fig:fig3_case2_summary}a), despite containing 36 rather than 30 years. An energy system that is optimised on a biased climate dataset is less prepared for the real climate and its inherent variability. \par

The performance differences between the compared selection methods reflect their fundamentally different search strategies for navigating the combinatorial explosion of possible subsets ($\binom{180}{30} \approx 10^{34}$ possible subsets for Case 1 and 2, $\binom{180}{5} \approx 10^{9}$ possible subsets for Case 3). SA proves to be the most efficient method for searching this large solution space, as it performs best across all three case studies. It works by iteratively swapping years in a candidate subset and accepting changes based on their effect on the SSWD score. Occasional acceptance of worsening swaps, with a probability that decreases over the run, allows SA to escape local minima while gradually converging to a near-optimal subset (Section~\ref{subsec:methods_simulated_annealing}).

K-Medoids clustering (Section~\ref{subsec:methods_k_medoids}) partitions the climate years into groups of similar years and selects the most central member of each group. This identifies distinct `types' of climate years, e.g. windy years, cold winters, high solar years, but does not directly optimise for distributional representativeness of the resulting subset. A set of cluster medoids can therefore miss important aspects of the full distribution, such as extremes. Despite this shortcoming, K-Medoids consistently performs second-best after simulated annealing (SA) when re-initialised many times and the subset with the lowest seasonal sliced Wasserstein distance (SSWD) is retained. \par

Filtered random search (Section~\ref{subsec:methods_frs}) extends random search (Section~\ref{subsec:methods_random_draws}) with sequential pre-filtering on cheap metrics, screening out poor candidates early so that more candidates can be evaluated within the same computational budget. Its design target is `not bad anywhere' rather than explicit optimisation of an aggregate score like the SSWD. Any candidate exceeding a threshold on any single variable is rejected, regardless of its overall representativeness. This filtering scales poorly with increasing dimensionality. Across cases, the pre-filtering in FRS degrades from a somewhat useful screening step at national scale (Case 1) into a near-total barrier at European scale (Cases 2 and 3), where it performs worse than plain random search in terms of optimising SSWD score. With many countries and variables, even a globally good subset may exceed a single per-variable threshold and be discarded. \par

The results above compare each method's best-found subset across 25 runs, each with many ($\sim 5000$) iterations. In practice, however, a naive user might execute only a single iteration of K-Medoids or (filtered) random search and adopt its output directly. For such single-draw usage, the median single-draw score (Tables~\ref{tab:tab1_case1_summary_table}--\ref{tab:tab3_case3_summary_table}) is the more relevant benchmark. Filtered random search (FRS) performs best on this metric, confirming that its pre-selection step adds value in filtering out the least representative subsets. K-Medoids offers a modest improvement over pure random draws by selecting medoid years of clusters. Even the best single-draw method, however, produces subsets far inferior to those obtained through repeated runs. SA has no meaningful single-iteration result, because it operates by iterative refinement. It is therefore excluded from this single-draw comparison. \par

The SSWD itself proves to be the appropriate evaluation metric for representativeness. The synthetic example cases in Supplementary Information~\ref{subsec:SI_SWD_illustrations} demonstrate that simpler statistical metrics fail in progressively more complex scenarios. The SWD captures both marginal distributions and inter-variable correlation structure (Figure~\ref{fig:S3_case3_swd_illustration}), and the seasonal extension (SSWD) prevents compensating errors across seasons (Figure~\ref{fig:S4_case4_swd_illustration}). This seasonal representation is essential for energy system applications, where the system implications of e.g.\ a low-wind period depend strongly on whether it occurs in winter or summer. \par

\subsection{Limitations}
The proposed methodology has three limitations. First, the selected subset is representative of the joint distribution defined by the six CMIP6 models and single realisations present in PECDv4.2. It may be less representative with respect to other equally valid model choices or different realisations of internal variability. This limitation is, however, inherent to all studies selecting representative periods from a finite reference dataset. Second, as with any heuristic optimisation, SA provides no guarantee that the global optimum has been found. The convergence behaviour (small spread across 25 independent runs and many iterations without found improvements within a run) suggests that solutions are near-optimal. The true optimality gap remains unknown because brute-force evaluation of all possible subsets is infeasible. Running SA multiple times and keeping the best result mitigates this uncertainty in practice. Third, representativeness in energy-meteorological space does not automatically imply optimality for usage in every energy system study \cite{vanderwielEnsembleClimateimpactModelling2020}. Studies focussing on specific extremes may still benefit from application-specific processing on top of our representative `base' subset. This can for example be done by modifying the cost function to weigh custom-chosen extremes more heavily. An SSWD-optimised subset nevertheless ensures that no significant climate-energy signal is systematically missing from the input, which is the minimum requirement for robust energy system analysis. \par

\subsection{Practical recommendations}
Based on this study, we provide five concrete recommendations for climate year selection in energy system studies:
\begin{enumerate}[label=(\roman*)]
    \item \textbf{Use simulated annealing for climate year selection.} SA achieves the lowest SSWD across 25 initialisations in all three cases, at roughly equal computational cost to K-Medoids and (filtered) random search (Tables~\ref{tab:tab1_case1_summary_table}--\ref{tab:tab3_case3_summary_table}, best overall score).
    \item \textbf{Use the SSWD to quantify representativeness.} The SSWD captures marginal distributions, inter-variable correlation structure, and seasonal balance in a single score (S.I. Section \ref{subsec:SI_SWD_illustrations}). Simpler metrics fail to capture the joint distribution structure, and an annual-only metric allows seasonal biases to compensate between winter and summer.
    \item \textbf{Run many iterations and keep the best result.} Any stochastic selection method (SA, K-Medoids, (filtered) random search) should be run for many iterations (here: $\sim 5000$), keeping the best result in terms of SSWD. Single-draw usage yields far-from-optimal subsets regardless of method.
    \item \textbf{Use the luck factor as a near-optimality diagnostic.} Re-initialising a run many times (here: 25 runs) is of secondary importance for SA, given its low dependence on luck. A low luck factor is, however, an indicator of near-optimality and therefore a useful diagnostic for new applications.
    \item \textbf{Define climate years as hydrological years (April 1st--March 31st).} April 1st is the start of the hydropower reservoir filling season, and starting the year in spring prevents breaking up resource adequacy stress events in winter (Section~\ref{subsec:methods_year_definition}).
\end{enumerate}

The SA+SSWD approach is recommended for selecting subsets of multiple complete climate years. Some detailed energy studies require even further data reduction, to sub-annual periods or aggregated time steps. For those applications, we recommend applying application-specific data reduction \textit{on top of} an SA-selected representative subset, not instead of it. Techniques such as time-series aggregation \cite{hoffmannReviewTimeSeries2020}, cost-driven clustering \cite{chatzistylianosAssessingImpactClimate2026}, or importance subsampling \cite{hilbersImportanceSubsamplingImproving2019} should take the SA-selected subset as their source dataset, ensuring that the starting point of any further reduction is in itself representative of the full climate.

\section{Conclusion}
\label{sec:conclusion}
SA optimised against the SSWD cost function consistently produces the most representative climate year subsets across spatial domains and subset sizes. SA subsets achieve a 3--15\,\% lower cost function value than the next-best method, and 14--26\,\% lower than the worst, depending on the case. They achieve an effective sample size of four to five times their actual size, and are roughly 2.5--3.5 times more representative than current ENTSO-E practice \cite{EuropeanResourceAdequacy}. \par

This study deliberately takes a bottom-up approach to climate year selection: optimising the representativeness of the climate-energy input dataset before any energy-study-specific processing occurs. This guarantees that the source dataset entering any energy system study is of validated and known representativeness in terms of climate variability, something purely energy-modelling focussed methods cannot ensure if their starting point is already biased. Those methods, such as time-series aggregation \cite{hoffmannReviewTimeSeries2020} or cost-driven clustering \cite{chatzistylianosAssessingImpactClimate2026} are typically not developed to bridge the gap between climate science (large ensembles) and energy system studies (a handful of years) \cite{craigOvercomingDisconnectEnergy2022}. They instead further reduce an already small, possibly unrepresentative set for the most detailed models. SA climate year selection optimised against the SSWD score closes that gap. A natural next step is to propagate SA-selected subsets through a power system model such as PyPSA \cite{brownPyPSAPythonPower2018}, to quantify whether the improved energy-climatological representativeness translates into measurable differences in capacity expansion decisions, system cost, and resource adequacy metrics. \par

The provided methodology is agnostic to the application domain. The same SA-with-SSWD procedure can be applied to any setting where a representative (multivariate) subset of complete years or other periods must be selected from a larger dataset. Possible applications include wind energy yield assessments \cite{pusatNewReferenceWind2021}, building energy simulations \cite{nikMakingEnergySimulation2016}, hydrological studies \cite{pechlivanidisInformationTheoryApproach2018}, and agricultural impact modelling \cite{ruaneSelectionRepresentativeSubset2017}. Energy system study outcomes are highly sensitive to climate data inputs \cite{rugglesPlanningReliableWind2024, grochowiczIntersectingNearoptimalSpaces2023, gotskeDesigningSectorcoupledEuropean2024, pecoraQuantifyingImpactsWeather2025}. Adopting this methodology can therefore substantially improve the robustness of investment planning and resource adequacy assessments at no additional computational cost, while potentially preventing billions of misinvested societal funds.


\section*{CRediT Author Statement}
Conceptualization: BvD, LPS;
Formal Analysis, Methodology, Investigation, and Visualization: BvD;
Writing - Original Draft: BvD;
Writing - Review \& Editing: \emph{All listed authors}; 
Funding Acquisition, Supervision: KvdW . 

\section*{Acknowledgments}
We thank the members of ET Climate from ENTSO-E \& TenneT EPN for helpful discussions. 

The content of this paper and the views expressed in it are solely the authors' responsibilities, and do not necessarily reflect the views of KNMI, RTE, and TenneT TSO.

\section*{Open research} 
The implementation of the analysis shown, all code used to generate the figures, and the methods to pre-process the data for the cases discussed as presented in this study are available at GitHub via \\ \url{https://github.com/bramvanduinen/weather_year_selection} with the MIT license. The GitHub-repository will become publicly available upon acceptance of the manuscript.

The dataset of PECDv4.2 containing the renewable resource capacity factors used in this study are available as part of the C3S Energy dataset. Copernicus Climate Change Service (2024): Climate and energy related variables from the Pan-European Climate Database derived from reanalysis and climate projections. Copernicus Climate Change Service (C3S) Climate Data Store (CDS). DOI: \url{https://doi.org/10.24381/cds.f323c5ec}

\renewcommand\refname{References}
\printbibliography 

\newpage

\appendix
\beginsupplement
\section*{Supplementary Information}

\section{Illustration of (seasonal) sliced-Wasserstein distance metric}
\label{subsec:SI_SWD_illustrations}
In this subsection, we establish the (seasonal) sliced-Wasserstein distance as the appropriate metric to measure representative subsets. We do this by showing several examples with synthetic data. In all examples, the sliced-Wasserstein distance (SWD, or SSWD in the case of seasonal SWD) clearly identifies the most-representative subset, where other metric might fall short. \par

\subsection{Example 1: Subsets with different mean}
Example 1 is the simplest. Here, subset A and subset B are both subsets of a larger reference set. The subset contains two (fictional) variables X and Y. Subset A captures the mean of the reference set well across both variables, subset B does not and is thus less representative in the most simple regard.

\noindent In this example, it is trivial to distinguish the representative subset (A) from the other not--representative one (B), both simple metrics like the mean absolute error, but also the more sophisticated SWD catch the better subset, see Table \ref{tab:S1_case1_swd_metrics}.

\begin{figure}[!h]
    \centering
    \includegraphics[width=0.85\linewidth]{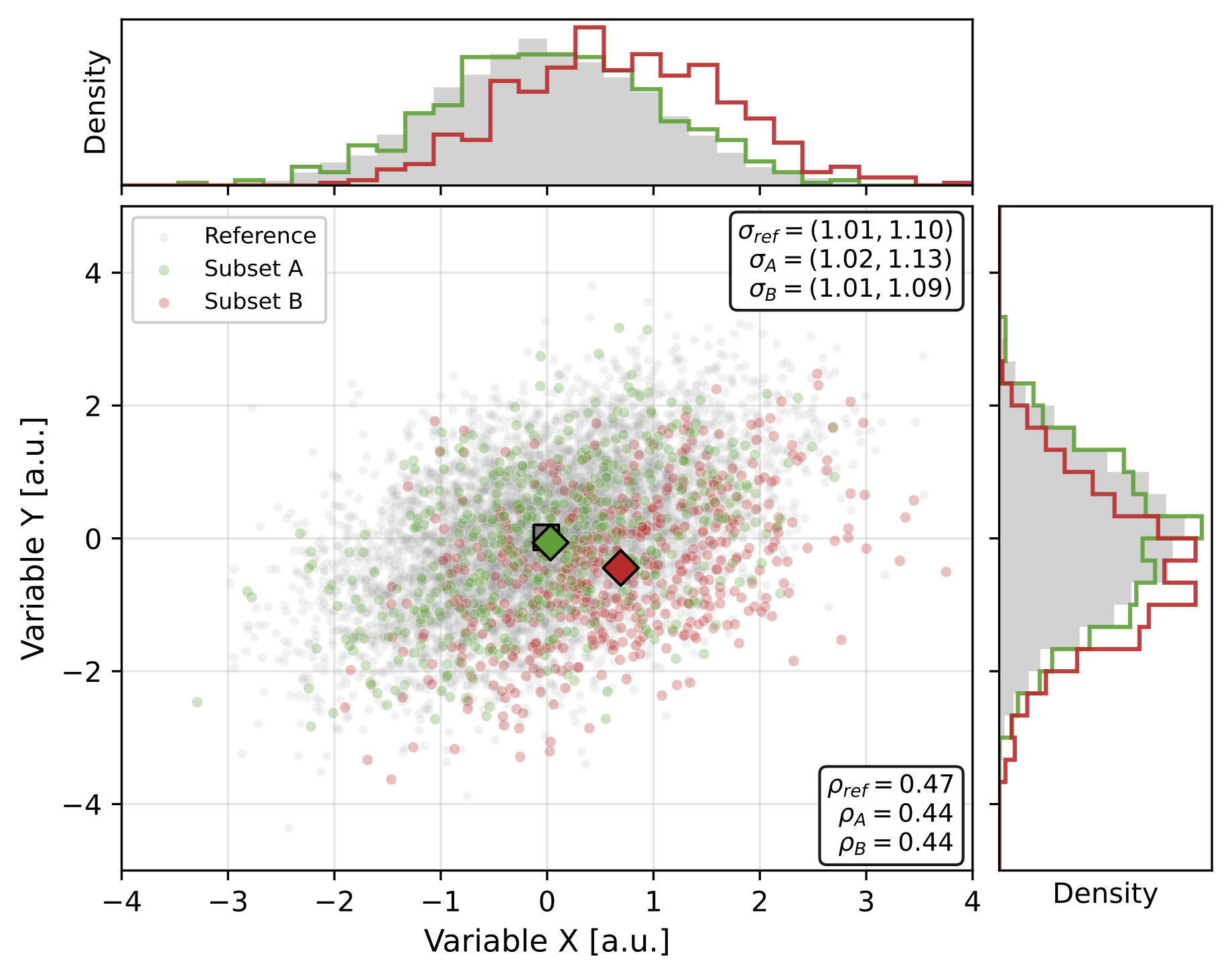}
    \caption{Example 1: Subsets with different means. The mean of the (sub)set is indicated by the diamonds. $\sigma$ gives the standard deviation of each set in variable ($X$, $Y$), $\rho$ gives the correlation between variables $X$ and $Y$ in each set.}
    \label{fig:S1_case1_swd_illustration}
\end{figure}

\begin{table}[hb]
\centering
\caption{Metrics for example 1 (Fig. \ref{fig:S1_case1_swd_illustration}), comparison of evaluation metrics for a good and bad subset. All metrics are calculated for the subset with respect to the reference set. The best score per metric is shown in bold. KS-statistic refers to the Kolmogorov-Smirnov statistic. QQ-MAE refers to the quantile-quantile mean absolute error. For univariate scores, the sum of the scores over all variables is taken.}
\begin{tabular}{l|cc}
\hline
\textbf{Metric} & \textbf{Subset A (good)} & \textbf{Subset B (bad)} \\
\hline
Mean absolute error of mean & \textbf{0.0588} & 0.5798 \\
KS statistic (sum over variables)     & \textbf{0.0880} & 0.4536 \\
QQ-MAE (sum over variables)           & \textbf{0.1723} & 1.1690 \\
Energy distance (sum over variables)  & \textbf{0.1066} & 0.8427 \\
SWD                         & \textbf{0.0985} & 0.5278 \\
\hline
\end{tabular}

\label{tab:S1_case1_swd_metrics}
\end{table}

\clearpage
\subsection{Example 2: Subsets with similar mean, but different variability}
Here, also a quite simple example. The subsets have a similarly representative mean, subset B is slightly better in terms of the mean per variable. However, subset B has too little variability compared to the reference. Therefore, subset A is more representative. The mean absolute error of the mean is not able to capture representativeness in terms of variability per variable. 

\begin{figure}[!h]
    \centering
    \includegraphics[width=0.85\linewidth]{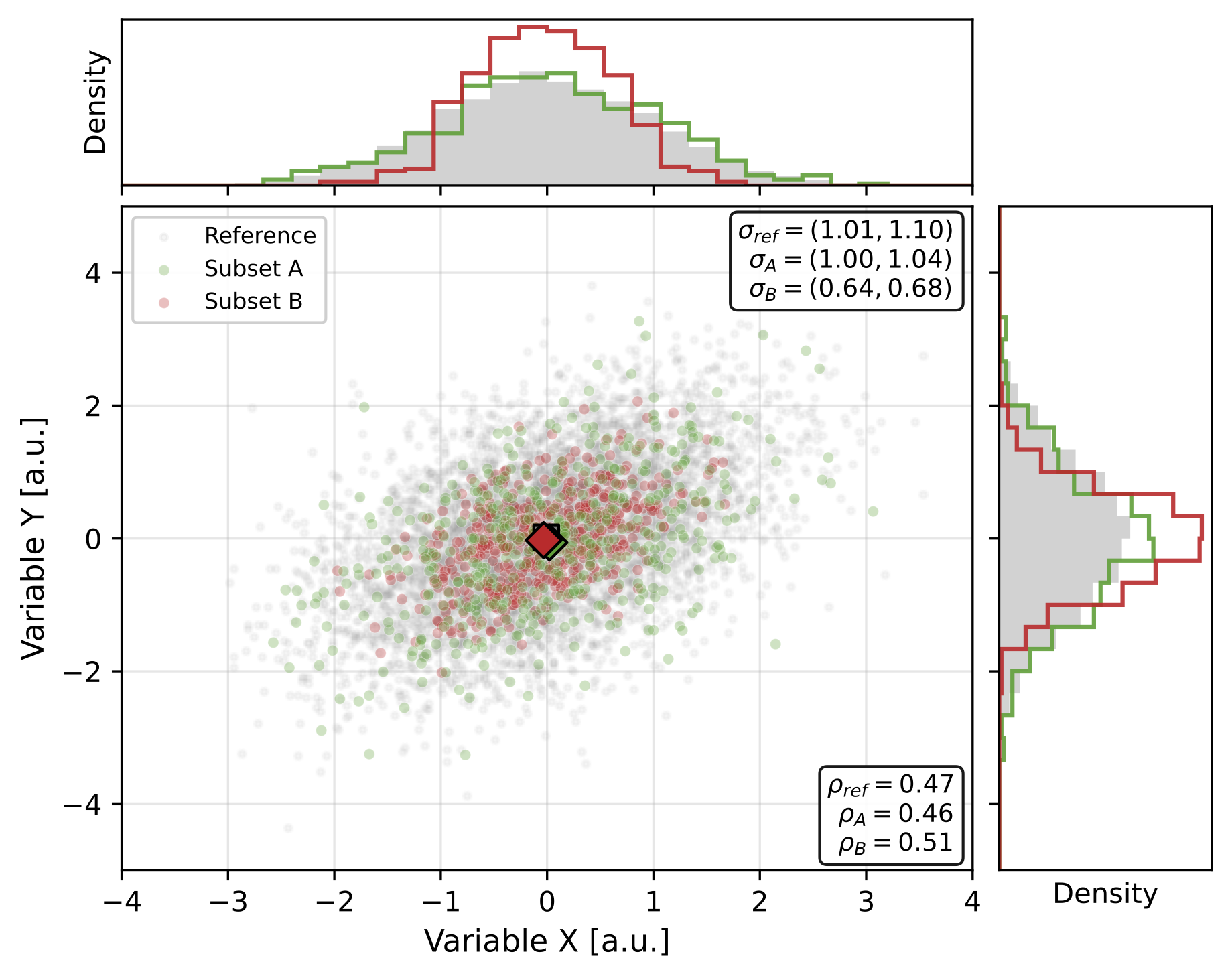}
    \caption{Example 2: Subsets with a similar mean, but different variability.}
    \label{fig:S2_case2_swd_illustration}
\end{figure}

\begin{table}[hb]
\centering
\caption{As Table \ref{tab:S1_case1_swd_metrics}, but here for example 2 (Fig. \ref{fig:S2_case2_swd_illustration}).}
\begin{tabular}{l|cc}
\hline
\textbf{Metric} & \textbf{Subset A (good)} & \textbf{Subset B (bad)} \\
\hline
Mean absolute error of mean & 0.0535 & \textbf{0.0329} \\
KS statistic (sum over variables)     & \textbf{0.1000} & 0.2688 \\
QQ-MAE (sum over variables)           & \textbf{0.1671} & 0.6857 \\
Energy distance (sum over variables)  & \textbf{0.1133} & 0.4605 \\
SWD                         & \textbf{0.0901} & 0.4068 \\
\hline
\end{tabular}
\label{tab:S2_case2_swd_metrics}
\end{table}

\clearpage
\subsection{Example 3: Subsets with similar univariate distribution, but different correlation.}
Here, a slightly more complex example. Subset B has slightly more representative univariate distributions. However, it does not have a representative correlation structure between the variables. Therefore, subset A is more representative. Only the SWD manages to identify this subset as the better subset.

\begin{figure}[!h]
    \centering
    \includegraphics[width=0.85\linewidth]{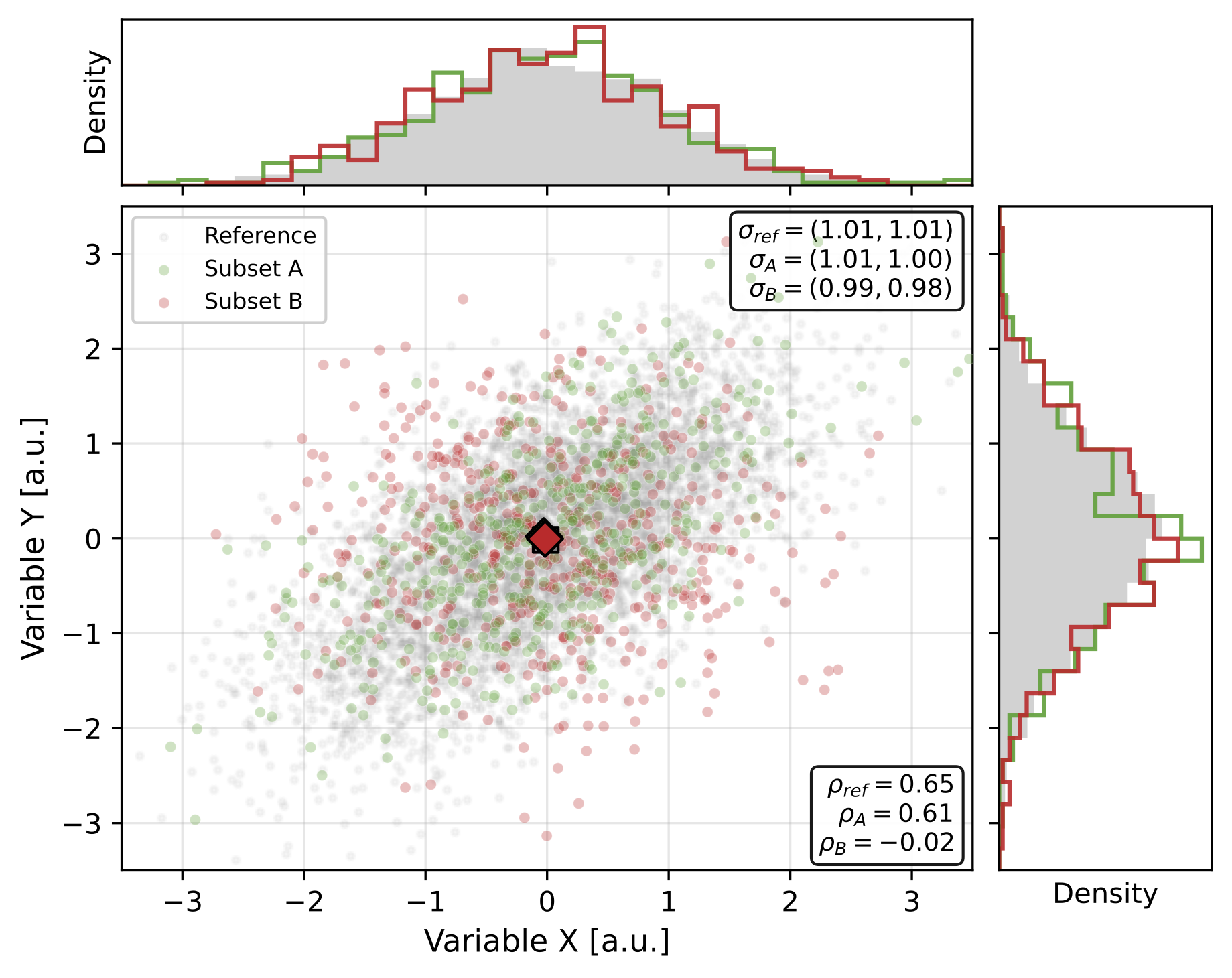}
    \caption{Example 3: Subsets with similar univariate distributions, but different correlation.}
    \label{fig:S3_case3_swd_illustration}
\end{figure}

\begin{table}[hb]
\centering
\caption{As Table \ref{tab:S1_case1_swd_metrics}, but here for example 3 (Fig. \ref{fig:S3_case3_swd_illustration}).}
\begin{tabular}{l|cc}
\hline
\textbf{Metric} & \textbf{Subset A (good)} & \textbf{Subset B (bad)} \\
\hline
Mean absolute error of mean & 0.0267          & \textbf{0.0081} \\
KS statistic (sum over variables)     & 0.0720          & \textbf{0.0552} \\
QQ-MAE (sum over variables)           & 0.1182          & \textbf{0.0929} \\
Energy distance (sum over variables)  & 0.0810          & \textbf{0.0546} \\
SWD                         & \textbf{0.0721} & 0.2443          \\
\hline
\end{tabular}
\label{tab:S3_case3_swd_metrics}
\end{table}

\clearpage
\subsection{Example 4: Subsets with similar annual joint-distribution, but different seasonal distribution}
Here, the most detailed example. Subset B has a slightly better representation year-round. However, there are compensating errors in the season. Therefore, on a seasonal basis, subset A is more representative. For energy system modelling, it is important to get seasonal representation. Therefore, subset A is more representative. Only the seasonal SWD (SSWD) manages to identify the better subset.

\begin{figure}[!h]
    \centering
    \includegraphics[width=\linewidth]{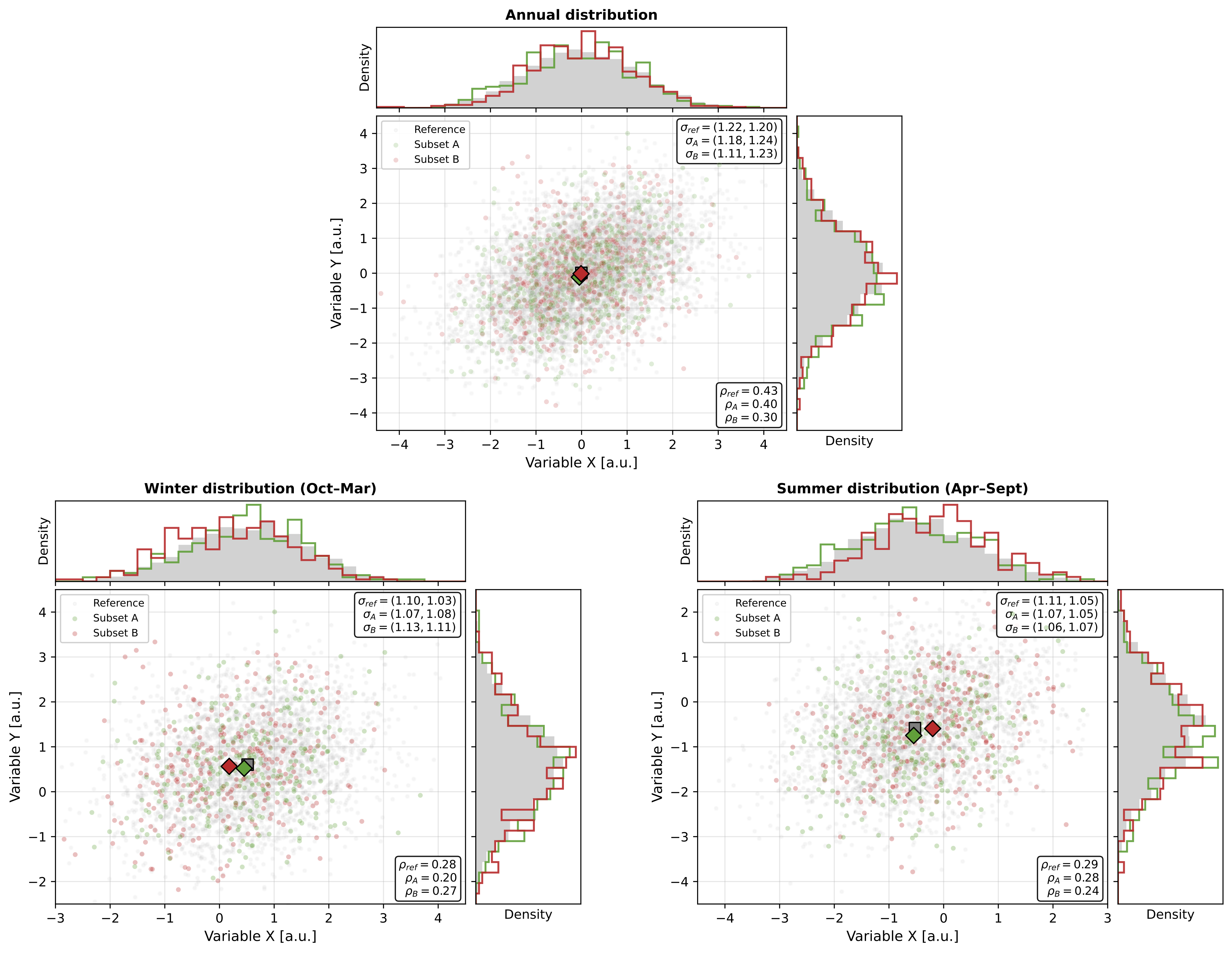}
    \caption{Example 4: Subsets with similar annual univariate distributions and correlation, but with different seasonal distributions.}
    \label{fig:S4_case4_swd_illustration}
\end{figure}

\begin{table}[hb!]
\centering
\caption{As Table \ref{tab:S1_case1_swd_metrics}, but here for example 4 (Fig. \ref{fig:S4_case4_swd_illustration}).}
\begin{tabular}{l|cc}
\hline
\textbf{Metric} & \textbf{Subset A (good)} & \textbf{Subset B (bad)} \\
\hline
Mean absolute error of mean & 0.0826          & \textbf{0.0146} \\
KS statistic (sum over variables)     & 0.1023          & \textbf{0.0748} \\
QQ-MAE (sum over variables)           & 0.2298          & \textbf{0.1781} \\
Energy distance (sum over variables)  & 0.1397          & \textbf{0.1034} \\
SWD (annual)                & 0.1146          & \textbf{0.1097} \\
SSWD (seasonal)             & \textbf{0.1475} & 0.2817          \\
\hline
\end{tabular}
\label{tab:S4_case4_swd_metrics}
\end{table}

\clearpage

\section{Selection algorithms}
\label{subsec:SI_selection_algorithms}
For each of the selection algorithms used for the analysis as described in Section~\ref{sec:methods} we give the pseudo-code here. For clarity they are shown on their own per page to not break the logic presented. 

\subsection{Random search}

\begin{algorithm}[H]
\caption{Random search}\label{alg:random_draw}
\KwIn{Reference dataset $\mathcal{D} = \{y_1, \dots, y_N\}$ of $N$ climate years, subset size $X$, number of repetitions $M$}
\KwOut{Subset $\mathcal{S}^* \subset \mathcal{D}$ of $X$ representative climate years and associated cost $c^*$}

Set $c^* \leftarrow \infty$\;

\For{$m \leftarrow 1$ \KwTo $M$}{
    Draw subset $\mathcal{S}^{(m)}$ by sampling $X$ years uniformly at random without replacement from $\mathcal{D}$\;
    Compute cost $c^{(m)} = \mathrm{SSWD}\!\left(\mathcal{S}^{(m)},\, \mathcal{D}\right)$\;
    \If{$c^{(m)} < c^*$}{
        $c^* \leftarrow c^{(m)}$;\quad $\mathcal{S}^* \leftarrow \mathcal{S}^{(m)}$\;
    }
}
\Return{$\mathcal{S}^*,\, c^*$}
\end{algorithm}

\clearpage
\subsection{Filtered Random Search (FRS
)}
\begin{algorithm}[H]
\caption{Filtered random search (FRS)}\label{alg:frs}
\KwIn{Reference dataset $\mathcal{D} = \{y_1, \dots, y_N\}$ of $N$ climate years, subset size $X$, number of iterations $M$, number of threshold-tuning iterations $M_{\mathrm{tune}}$, threshold percentile $p$, set of production and demand variables $\mathcal{V}$}
\KwOut{Subset $\mathcal{S}^* \subset \mathcal{D}$ of $X$ representative climate years and associated cost $c^*$}
\BlankLine
\textbf{Threshold tuning}
\BlankLine
\For{$m \leftarrow 1$ \KwTo $M_{\mathrm{tune}}$}{
    Draw subset $\mathcal{S}^{(m)}$ by sampling $X$ years uniformly at random without replacement from $\mathcal{D}$\;
    \ForEach{$v \in \mathcal{V}$}{
        \ForEach{$t \in \{w, s\}$, \textnormal{(winter, summer)}}{
            {Compute $d_{\mathrm{mean}}(\mathcal{S}^{(m)}, \mathcal{D}, v, t)$}\;
            {Compute $d_{\mathrm{energy}}(\mathcal{S}^{(m)}, \mathcal{D}, v, t)$\;}
        }
    }
}
Set thresholds $\tau_{\mathrm{mean},v,t} = \mathrm{P}_p\!\left(\left\{d_{\mathrm{mean}}(\mathcal{S}^{(m)}, \mathcal{D}, v, t)\right\}_{m=1}^{M_{\mathrm{tune}}}\right)$ and $\tau_{\mathrm{energy},v,t} = \mathrm{P}_p\!\left(\left\{d_{\mathrm{energy}}(\mathcal{S}^{(m)}, \mathcal{D}, v, t)\right\}_{m=1}^{M_{\mathrm{tune}}}\right)$ for each $v \in \mathcal{V}$, $t \in \{w, s\}$, and $\mathrm{P}_p$ is the $p$-th percentile of the set of values;\
\BlankLine
\textbf{Main loop}\\
\BlankLine
Set $c^* \leftarrow \infty$\;
\For{$m \leftarrow 1$ \KwTo $M$}{
    Draw subset $\mathcal{S}^{(m)}$ by sampling $X$ years uniformly at random without replacement from $\mathcal{D}$\;
    \ForEach {$v \in \mathcal{V}$}{
        \ForEach{$t \in \{w, s\}$}{
            Compute $d_{\mathrm{mean}}(\mathcal{S}^{(m)}, \mathcal{D}, v, t)$\;
            \lIf{$d_{\mathrm{mean}}(\mathcal{S}^{(m)}, \mathcal{D}, v, t) > \tau_{\mathrm{mean},v,t}$}{\textbf{discard} $\mathcal{S}^{(m)}$ and \textbf{continue} to next $m$}
            Compute $d_{\mathrm{energy}}(\mathcal{S}^{(m)}, \mathcal{D}, v, t)$\;
            \lIf{$d_{\mathrm{energy}}(\mathcal{S}^{(m)}, \mathcal{D}, v, t) > \tau_{\mathrm{energy},v,t}$}{\textbf{discard} $\mathcal{S}^{(m)}$ and \textbf{continue} to next $m$}
        }
    }
    Compute cost $c^{(m)} = \mathrm{SSWD}\!\left(\mathcal{S}^{(m)},\, \mathcal{D}\right)$\;
    \If{$c^{(m)} < c^*$}{
        $c^* \leftarrow c^{(m)}$;\quad $\mathcal{S}^* \leftarrow \mathcal{S}^{(m)}$\;
    }
}
\Return{$\mathcal{S}^*,\, c^*$}
\end{algorithm}
The threshold-percentile is set separately per case, to ensure that sufficient but not too much (order 1-in-1000) candidate subsets pass the pre-filtering steps. For Case 1 (single country, 30~years) $p = 25$, for Case 2 and 3 (whole Europe, 30~years and 5~years) $p = 85$. 

\clearpage
\subsection{K-Medoids}

\begin{algorithm}[H]
\caption{K-Medoids clustering}\label{alg:k_medoids}
\KwIn{Reference dataset $\mathcal{D} = \{y_1, \dots, y_N\}$ of $N$ climate years, subset size $X$, number of repetitions $M$}
\KwOut{Subset $\mathcal{S}^* \subset \mathcal{D}$ of $X$ representative climate years and associated cost $c^*$}

Compute feature vector $\mathbf{f}_i = [\mu_1^{w}(y_i),\, \sigma_1^{w}(y_i),\, \mu_1^{s}(y_i),\, \sigma_1^{s}(y_i),\, \dots,\, \mu_V^{w}(y_i),\, \sigma_V^{w}(y_i),\, \mu_V^{s}(y_i),\, \sigma_V^{s}(y_i)] \in \mathbb{R}^{4V}$ for each $y_i \in \mathcal{D}$, where $\mu_v^{w}$, $\sigma_v^{w}$ and $\mu_v^{s}$, $\sigma_v^{s}$ denote the mean and standard deviation of variable $v$ in the extended winter (October--March) and summer (April--September) seasons respectively, and $V$ is the total number of production and demand variables\;

Set $c^* \leftarrow \infty$\;

\For{$m \leftarrow 1$ \KwTo $M$}{
    Initialise $X$ cluster centroids via K-Medoids$++$ initialisation\;
    Iteratively reassign each $y_i$ to its nearest centroid and update centroids until convergence, yielding clusters $\mathcal{C}_1^{(m)}, \dots, \mathcal{C}_X^{(m)}$\;
    Form candidate subset $\mathcal{S}^{(m)} = \left\{ \operatorname{medoid}(\mathcal{C}_k^{(m)}) \right\}_{k=1}^X$, where $\operatorname{medoid}(\mathcal{C}_k) = \arg\min_{y \in \mathcal{C}_k} \sum_{y' \in \mathcal{C}_k} \lVert \mathbf{f}_{y} - \mathbf{f}_{y'} \rVert_2$, with $y$ the candidate medoid year and $y'$ ranging over all other years in the cluster\;
    Compute cost $c^{(m)} = \mathrm{SSWD}\!\left(\mathcal{S}^{(m)},\, \mathcal{D}\right)$\;
    \If{$c^{(m)} < c^*$}{
        $c^* \leftarrow c^{(m)}$;\quad $\mathcal{S}^* \leftarrow \mathcal{S}^{(m)}$\;
    }
}
\Return{$\mathcal{S}^*,\, c^*$}
\end{algorithm}

\clearpage
\subsection{Simulated Annealing}

\begin{algorithm}[H]
\caption{Simulated annealing (SA)}\label{alg:sa}
\KwIn{Reference dataset $\mathcal{D} = \{y_1, \dots, y_N\}$ of $N$ climate years, subset size $X$, initial temperature $T_{\mathrm{start}}$, terminal temperature $T_{\mathrm{end}}$, cooling factor $c_r \in (0,1)$, reheating threshold $n_{\mathrm{stall}}$, maximum iterations $it_{\mathrm{max}}$}
\KwOut{Subset $\mathcal{S}^* \subset \mathcal{D}$ of $X$ representative climate years and associated cost $c^*$, number of iterations $it$}
Initialise $\mathcal{S}$ by sampling $X$ years uniformly at random without replacement from $\mathcal{D}$\;
Set $c \leftarrow \mathrm{SSWD}(\mathcal{S}, \mathcal{D})$;\quad $\mathcal{S}^* \leftarrow \mathcal{S}$;\quad $c^* \leftarrow c$;\quad $T \leftarrow T_{\mathrm{start}}$;\quad $n_{\mathrm{no\,imp}} \leftarrow 0$;\quad $it \leftarrow 0$\;
\While{$T > T_{\mathrm{end}}$ \textbf{and} $it < it_{\mathrm{max}}$}{
    Construct neighbour subset $\mathcal{S}'$ of $\mathcal{S}$ by swapping $k$ years for random replacement years (see Eq. \eqref{eq:replacement_small_subset}, \eqref{eq:replacement_large_subset})\;
    Compute $c' = \mathrm{SSWD}(\mathcal{S}', \mathcal{D})$\;
    \If{$c' < c$ \textbf{or} $\mathrm{random()} < \exp(-(c'-c)/T)$}{
        $\mathcal{S} \leftarrow \mathcal{S}'$;\quad $c \leftarrow c'$\;
    }
    \If{$c < c^*$}{
        $c^* \leftarrow c$;\quad $\mathcal{S}^* \leftarrow \mathcal{S}$;\quad $n_{\mathrm{no\,imp}} \leftarrow 0$\;
    }\Else{
        $n_{\mathrm{no\,imp}} \leftarrow n_{\mathrm{no\,imp}} + 1$\;
    }
    \If{$n_{\mathrm{no\,imp}} \geq n_{\mathrm{stall}}$}{
        $T \leftarrow 1.8 \cdot T$;\quad $n_{\mathrm{no\,imp}} \leftarrow 0$\;
    }
    $T \leftarrow c_r \cdot T$;\quad $it \leftarrow it + 1$\;
}
\Return{$\mathcal{S}^*,\, c^*, it$}
\end{algorithm}

The number of $k$ years to swap is determined stochastically, and depends on the number of representative climate years looked for. \par
For small subsets ($X < 10$):
\begin{equation}
\label{eq:replacement_small_subset}
    k = \begin{cases} 1 & \text{with probability } 0.75 \\ \mathcal{U}\{2, 3\} & \text{with probability } 0.25 \end{cases}
\end{equation}
\noindent For large subsets ($X > 10$):
\begin{equation}
\label{eq:replacement_large_subset}
    k = \begin{cases} \mathcal{U}\{1, 2, 3\} & \text{with probability } 0.50 \\ \mathcal{U}\{4, 5, 6\} & \text{with probability } 0.35 \\ \mathcal{U}\{7, 8, 9, 10\} & \text{with probability } 0.15 \end{cases}
\end{equation}
where $\mathcal{U}\{a, \dots, b\}$ denotes a discrete uniform draw from the integers $a$ through $b$. This scheme ensures that for small subsets a single swap is preferred, preserving most of the current solution. For large subsets more aggressive perturbations are also permitted, to explore a wider region of the solution space.

\clearpage

\section{Uncertainty in (seasonal) sliced-Wasserstein distance metric}
\begin{figure}[h!]
    \centering
    \includegraphics[width=\linewidth]{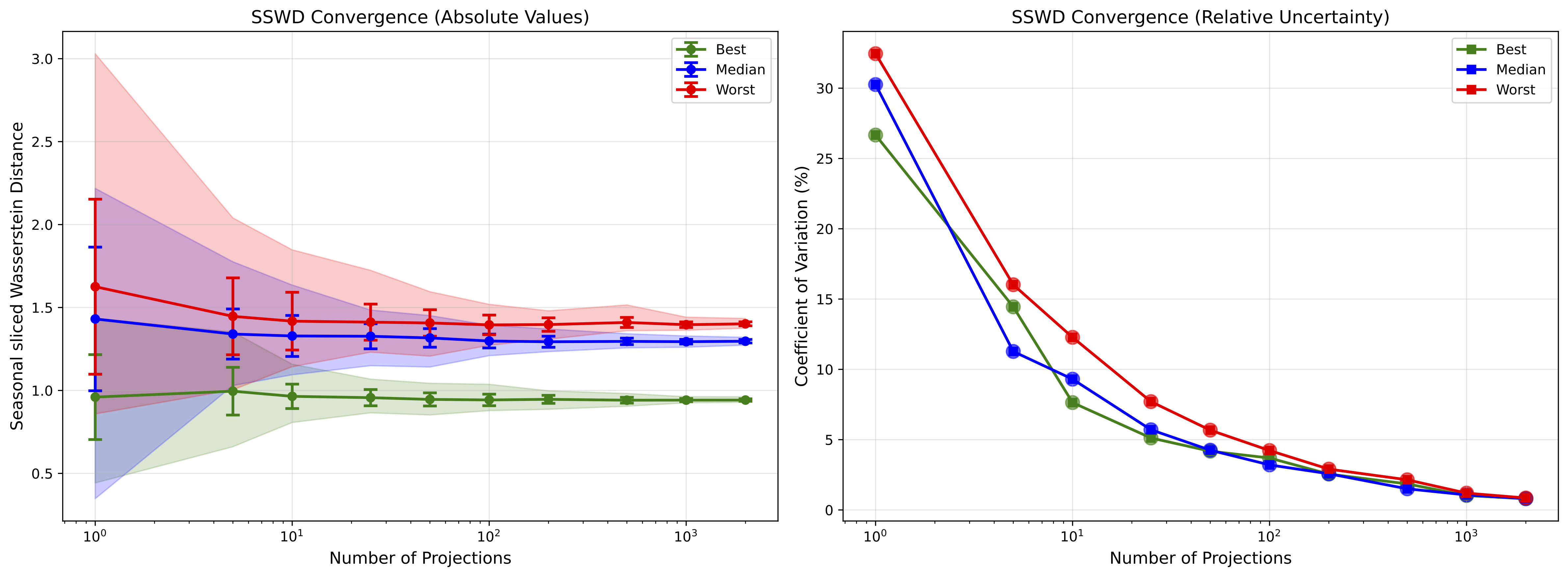}
    \caption{Uncertainty in SSWD as a function of the number of projections, for the Netherlands 30-year case (Case 1). Results are shown for the best, median, and worst year selections by SSWD score, based on 50 repeated calculations per projection count with different random seeds. Left: absolute SSWD values; error bars indicate the standard deviation and shading the full spread. Right: coefficient of variation ($CV = \sigma/\mu$).}
    \label{fig:S5_sswd_uncertainty_analysis_NL30y}
\end{figure}

\begin{figure}[h!]
    \centering
    \includegraphics[width=\linewidth]{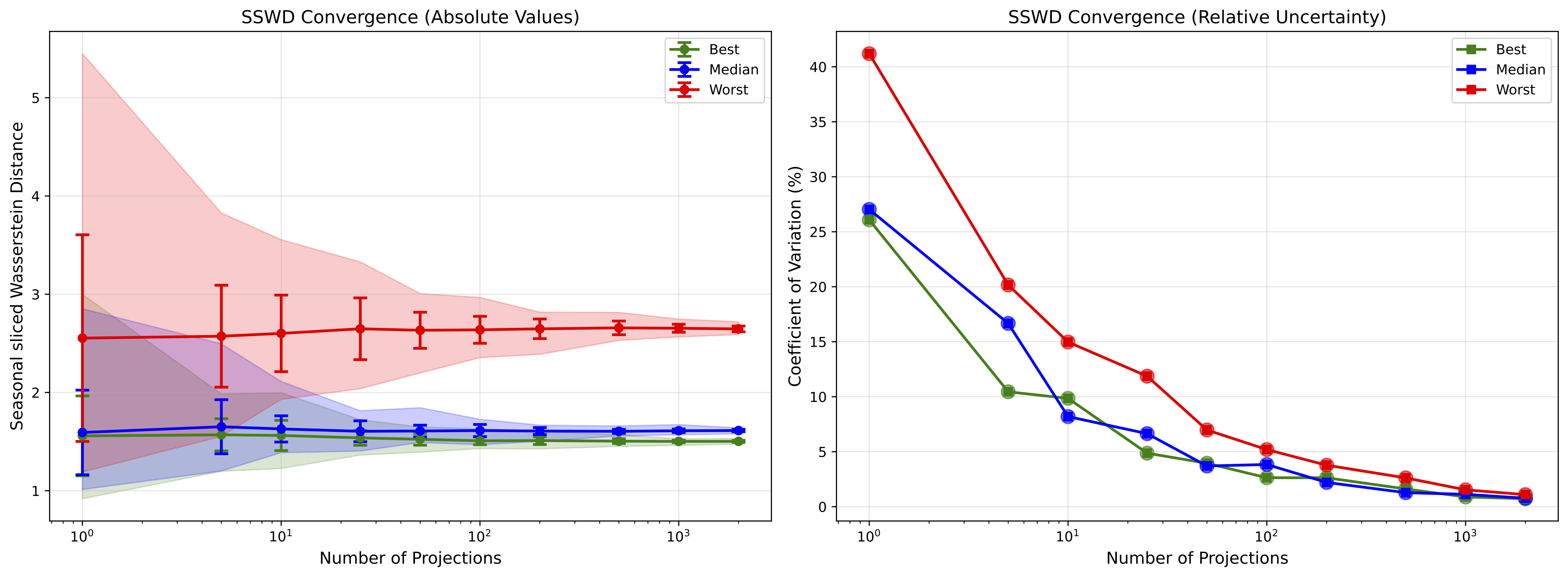}
    \caption{As Figure \ref{fig:S5_sswd_uncertainty_analysis_NL30y}, but here for the Europe 30-year case (Case 2).}
    \label{fig:S6_sswd_uncertainty_analysis_EU30y}
\end{figure}

\begin{figure}[h!]
    \centering
    \includegraphics[width=\linewidth]{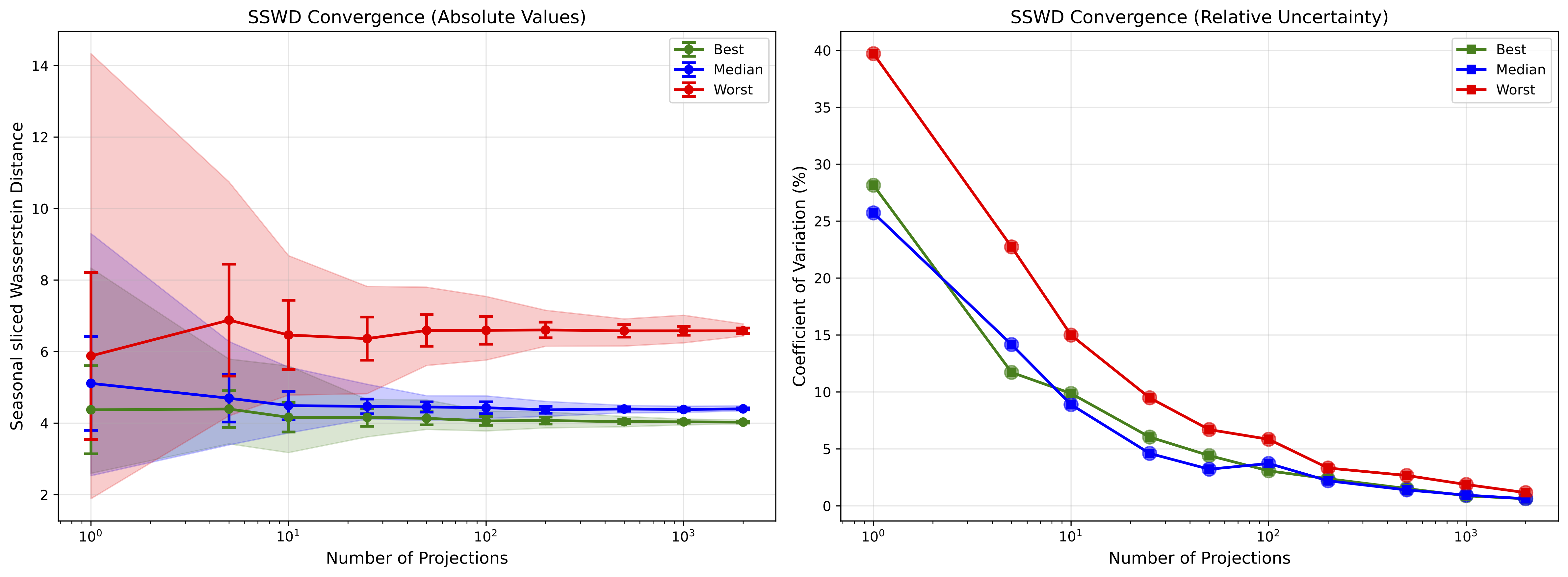}
    \caption{As Figure \ref{fig:S5_sswd_uncertainty_analysis_NL30y}, but here for the Europe 5-year case (Case 3).}
    \label{fig:S7_sswd_uncertainty_analysis_EU5y}
\end{figure}

As discussed in Section \ref{subsec:methods_costfunction}, due to the random nature of the projections, the (S)SWD carries some uncertainty that decreases with increasing $N_p$. Here, we illustrate this by calculating the SSWD based on 50 repeated calculations per projection count with different random seeds. It can be seen in Figure~\ref{fig:S5_sswd_uncertainty_analysis_NL30y}--\ref{fig:S7_sswd_uncertainty_analysis_EU5y} that the SSWD value converges per dataset with increasing number of projections. To balance computational efficiency and accuracy, we use $N_p = 50$ projections for initial search phase, and re-calculate with $N_p = 200$ projections for the 10~\% most promising subsets per run for each method. This results in a coefficient of variation of about 2--3 \%.


\clearpage
\section{Supplementary Figures}
\label{subsec:SI_supplementary_figures}
Additional figures of the results shown in Section~\ref{sec:results} are provided here. For the analysis see the main manuscript.

\begin{figure}[ht]
    \centering
    \includegraphics[width=0.7\linewidth]{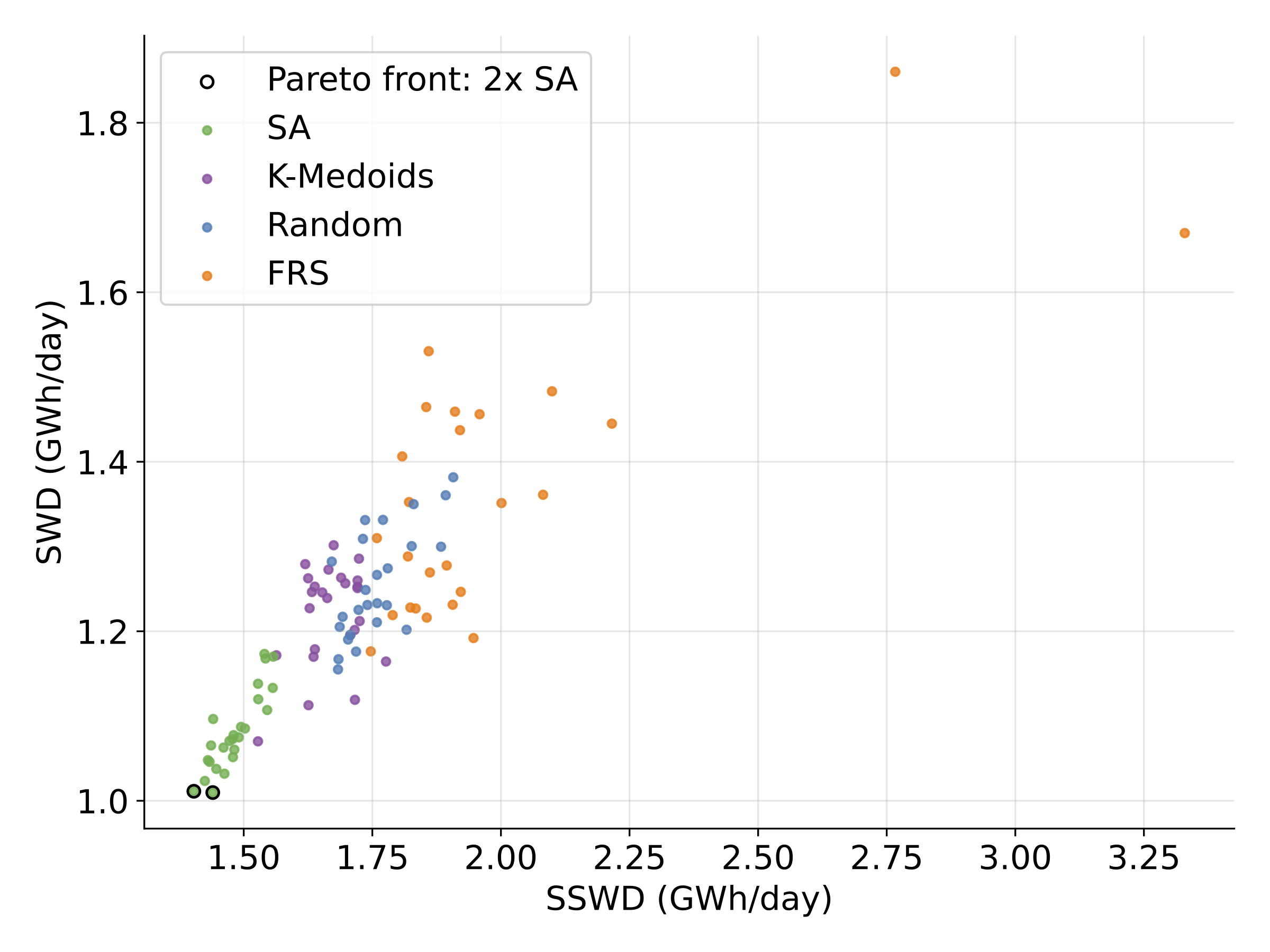}
    \caption{As Figure \ref{fig:fig2_case1_pareto}, but here for $X = 30$ years selected for Europe (Case 2). The Pareto front consists of two SA runs.}
    \label{fig:figS7_case2_pareto}
\end{figure}

\begin{figure}[hb]
    \centering
    \includegraphics[width=0.7\linewidth]{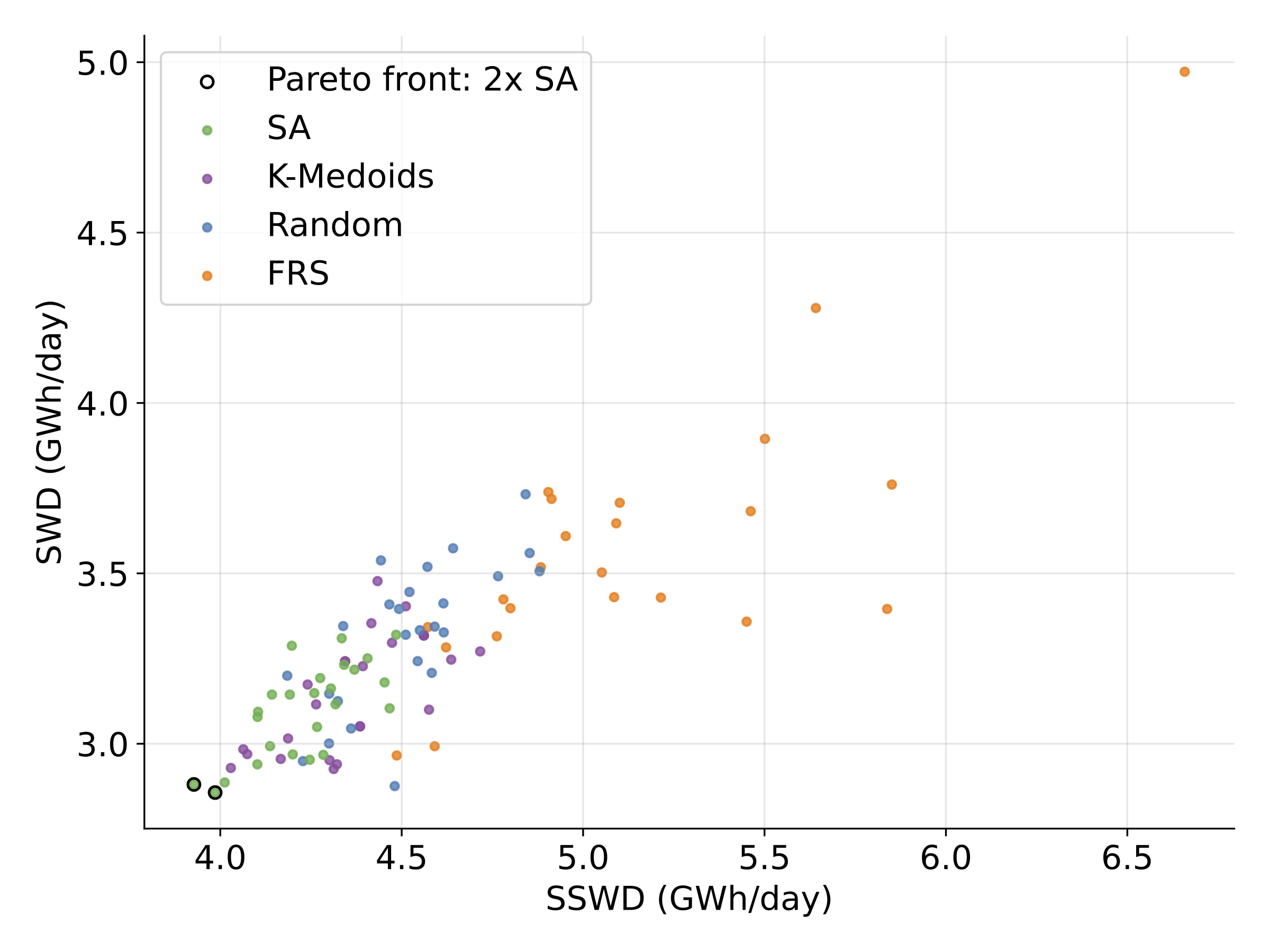}
    \caption{As Figure \ref{fig:fig2_case1_pareto}, but here for $X = 5$ years selected for Europe (Case 3). The Pareto front consists of two SA runs.}
    \label{fig:figS8_case3_pareto}
\end{figure}

\end{document}